\documentclass[twocol]{ametsoc}
\usepackage{dsfont}

\journal{jas}

\bibpunct{(}{)}{;}{a}{}{,}

\title{The linear response function of an idealized atmosphere. Part 1: \\ Construction using Green's functions and applications}

    \authors{Pedram Hassanzadeh\correspondingauthor{Pedram Hassanzadeh, 24 Oxford Street, Cambridge, MA 02138.}}
    \affiliation{Department of Earth and Planetary Sciences and Center for the Environment, Harvard University, Cambridge, Massachusetts}
    \email{hassanzadeh@fas.harvard.edu}

    \extraauthor{Zhiming Kuang}
    \extraaffil{\small Department of Earth and Planetary Sciences and John A. Paulson School of Engineering and Applied Sciences, Harvard University, Cambridge, Massachusetts}

\abstract{A linear response function (LRF) determines the mean-response of a nonlinear climate system to weak imposed forcings, and an eddy flux matrix (EFM) determines the eddy momentum and heat flux responses to mean-flow changes. Neither LRF nor EFM can be calculated from first principles due to the lack of a complete theory for turbulent eddies. Here the LRF and EFM for an idealized dry atmosphere are computed by applying numerous localized weak forcings, one at a time, to a GCM with Held-Suarez physics and calculating the mean-responses. The LRF and EFM for zonally-averaged responses are then constructed using these forcings and responses through matrix inversion. Tests demonstrate that LRF and EFM are fairly accurate. Spectral analysis of the LRF shows that the most excitable dynamical mode, the neutral vector, strongly resembles the model's Annular Mode. The framework described here can be employed to compute the LRF/EFM for zonally-asymmetric responses and more complex GCMs. The potential applications of the LRF/EFM constructed here are i) forcing a specified mean-flow for hypothesis-testing, ii) isolating/quantifying the eddy-feedbacks in complex eddy-mean flow interaction problems, and iii) evaluating/improving more generally-applicable methods currently used to construct LRFs or diagnose eddy-feedbacks in comprehensive GCMs or observations. As an example for iii, in Part 2, the LRF is also computed using the fluctuation-dissipation theorem (FDT), and the previously-calculated LRF is exploited to investigate why FDT performs poorly in some cases. It is shown that dimension-reduction using leading EOFs, which is commonly used to construct LRFs from the FDT, can significantly degrade the accuracy due to the non-normality of the operator.} 

\begin{document}

%% Necessary!
\maketitle

%%%%%%%%%%%%%%%%%%%%%%%%%%%%%%%%%%%%%%%%%%%%%%%%%%%%%%%%%%%%%%%%%%%%%
% MAIN BODY OF PAPER
%%%%%%%%%%%%%%%%%%%%%%%%%%%%%%%%%%%%%%%%%%%%%%%%%%%%%%%%%%%%%%%%%%%%%
%
\section{Introduction}\label{sec:intro}
How the different components of the climate system respond to imposed forcings is of significant importance for understanding the internal climate variability and anthropogenic climate change. For an external forcing $\mathrm{\mathbf{f}}$ that is weak enough so that the response of the nonlinear climate system varies linearly with $\mathrm{\mathbf{f}}$, the problem can be formulated as \citep{palmer99}
\begin{eqnarray}
\dot{\mathrm{\mathbf{x}}} = \pmb{\mathsf{L}} \, \mathrm{\mathbf{x}} + \mathrm{\mathbf{f}}
\label{eq:lrf0}
\end{eqnarray}
where $\mathrm{\mathbf{x}}$ is the state-vector response, i.e., deviation from the time-mean flow (\textit{mean-flow} hereafter) of the unforced ($\mathrm{\mathbf{f}}=0$) system, and $\pmb{\mathsf{L}}$ is the linear response function (LRF) of the system (see section~\ref{sec:eqs} for details). If $\pmb{\mathsf{L}}$ is known, we can immediately calculate not only the responses to various scenarios of external forcings (which can also be obtained from usually-expensive GCM simulations), but we can also find the forcing needed to achieve a prescribed response (the inverse problem), the most effective forcing (i.e., forcing producing the largest response), and the most excitable dynamical mode, i.e., the so-called \textit{neutral vector}, which is important for both internal variability and forced response \citep{marshall1993toward,goodman2002using}. 
%In this paper we focus on zonally-averaged forcings $\overline{\mathrm{\mathbf{f}}}$ and responses $\overline{\mathrm{\mathbf{x}}}$ in the atmosphere, and will continue to use $\pmb{\mathsf{M}}$ to refer to the LRF. Hence
%\begin{eqnarray}
%\dot{\overline{\mathrm{\mathbf{x}}}} = \pmb{\mathsf{M}} \, \overline{\mathrm{\mathbf{x}}} + \overline{\mathrm{\mathbf{f}}}.
%\label{eq:lrf1}
%\end{eqnarray}
%Overbears denote zonal-means hereafter.

Another problem of great interest is how eddies respond to changes in the mean-flow (which here refers to the flow averaged over an appropriate timescale much longer than the eddy timescale). Focusing on synoptic eddies in the atmosphere, for instance, the increase of eddy phase-speed in response to strengthening of the lower stratospheric winds \citep{chen2007phase} and the increase of eddy length-scale in response to changes in the atmosphere's thermal structure \citep{kidston11,riviere2011dynamical} have been suggested to cause the observed and projected poleward shift of the midlatitude jets. As another example, changes in the eddy momentum and heat fluxes in response to changes in the tropospheric zonal-wind and temperature play a critical role in the dynamics of the leading pattern of variability in the midlatitudes, i.e., the Northern and Southern Annular Modes \citep{robinson2000baroclinic,lorenz2001eddy}, while the tropospheric eddy flux response to stratospheric changes are important for the tropospheric-stratospheric coupling \citep{haynes1991downward,kushner2004stratosphere,song2004dynamical}. 
%Note that $u'$, $v'$, and $T'$, are the anomalous zonal and meridional velocities and temperature, respectively, and $\langle \cdot \rangle$ denote some appropriate time-averaging. 
The eddies modify the mean-flow through the eddy momentum and heat fluxes, which can be quantified, for example, using the Eliassen-Palm flux \citep{edmon1980eliassen}. The mean-flow also modifies the eddies. What complicates the eddy-mean flow interaction problems, not only in the atmosphere but also in other turbulent geophysical flows \citep[][chs. 7-10]{vallis2006atmospheric}, is that, despite extensive efforts, how eddy fluxes respond to a change in the mean-flow is not fully understood and a complete theory for this response is currently unavailable \citep{heldgfd,schneider2006general}. For example, as described in section~\ref{sec:eqs}, for the zonal-mean response in an idealized dry atmosphere (which is the focus of this paper), the state-vector consists of $\overline{u}$ and $\overline{T}$, and the problem can be formulated as        
\begin{equation}
\begin{bmatrix}
    \langle \overline{u'v'} \rangle \\
    \langle \overline{v'T'} \rangle
  \end{bmatrix}
=
\pmb{\mathsf{E}} \begin{bmatrix}
    \langle \overline{u} \rangle \\
    \langle \overline{T} \rangle
  \end{bmatrix}
\label{eq:efm}
\end{equation}
where the overbars and $\langle \cdot \rangle$ denote, respectively, the zonal-means and time-means; $\overline{u}$ and $\overline{T}$ are the responses in zonal-mean zonal-wind and temperature; and $\langle \overline{u'v'} \rangle$ and $\langle \overline{v'T'} \rangle$ are the responses of eddy momentum and heat fluxes to $\langle \overline{u} \rangle$ and $\langle \overline{T} \rangle$. In this paper we refer to $\pmb{\mathsf{E}}$ as the eddy flux matrix (EFM), which with the current state of understanding cannot be determined from the first principles. 
%The lack of a complete theory for eddies (and the resulting unavailability of $\pmb{\mathsf{E}}$ and similar operators) obscures a full understanding of various dynamical mechanisms and in particular the direction of causation, and complicates the interpretation and analysis of the observational data and outputs of comprehensive GCMs.            

The two issues discussed above are related: a major difficulty in calculating the LRFs is to accurately account for the eddy-feedbacks (because $\pmb{\mathsf{E}}$ is unknown), and as a result even in very simple models of the climate system, finding the LRFs remains a challenge. For example, in the Held-Suarez benchmark setup \citep{held1994proposal}, where the focus is on a dry atmosphere with simply zonally-symmetric boundary conditions and parametrization of radiation and planetary boundary-layer with, respectively, Newtonian cooling and Rayleigh drag, the LRF consists of four components: mean-flow advection, surface friction, Newtonian relaxation, and eddy-feedbacks. While in this setup the first three components are known analytically for a given mean-flow, the eddy-feedback is not, which renders the LRF indeterminable analytically.                   

The three common approaches to finding LRFs involve neglecting the eddy-feedbacks, parameterizing the eddy-feedbacks, or employing the fluctuation-dissipation theorem (FDT). The first approach is similar to the hydrodynamic linear stability analysis where the equations of motion are linearized around a mean-flow and all nonlinear terms (which contain eddy fluxes) are ignored \citep[][ch. 6]{vallis2006atmospheric}. Such an approach has provided valuable insight into some aspects of the atmospheric circulation \citep{hoskins1981steady,marshall1993toward,goodman2002using}, but because of the lack of eddy-feedbacks, can result in inaccurate LRFs \citep[e.g.,][]{branstator1998empirical} and LRFs with linearly unstable modes, which suggests, from Eq.~(\ref{eq:lrf0}), unbounded growth of $\mathrm{\mathbf{x}}$. Accounting for the eddy-feedbacks using diffusive \citep[e.g.,][ch. 10]{pavan1996diffusive,lapeyre2003diffusivity,vallis2006atmospheric} or stochastic \citep[e.g.,][]{farrell1996generalized,farrell1996generalizedII,farrell2003structural,zhang1999linear} closures might improve the accuracy of the LRF, but there are challenges associated with using these parameterizations in complex models, see, e.g., the reviews by \citet{franzke2015stochastic} and \citet{delsole2004stochastic}. 

The FDT, a powerful tool in statistical physics \citep{nyquist1928thermal,kubo1966fluctuation}, suggest that the LRF can be built for a GCM from lag-covariances of a long unforced simulation and even for the climate system itself from observations \citep{leith1975climate}. In principle, the FDT can produce a LRF that accurately takes into account the eddy-feedbacks and any other physical process that is present in the model or nature without the need for a detailed understanding of these processes. However, there are uncertainties in applicability of the FDT to the climate system, and testing FDT in GCMs \citep[e.g.,][]{gritsun2007climate,ring08,fuchs2015exploration} and simple models of geophysical turbulence \citep[e.g.,][]{majda2010low,lutsko2015applying} have produced mixed results. The FDT is further discussed in Part 2 of this study \citep{pedram16}.  

In this paper (Part 1) we take a different approach to finding the LRF (with accurate eddy-feedbacks) and EFM of a relatively simple GCM. We use Green's functions following the framework developed in \citet{kuang2010linear} to calculate the LRF for a cloud-system-resolving model, which has been used to study convectively-coupled waves \citep{kuang2010linear}, Walker cells \citep{kuang2012weakly}, and convective parameterization schemes \citep{nie2012responses,herman2013linear}. With problems in large-scale atmospheric circulation in mind, here we focus on the zonally-averaged forcings and responses of an idealized dry GCM, where the state-vector consists only of $\overline{u}$ and $\overline{T}$ (see section~\ref{sec:eqs} for details). The procedure to find the LRF and EFM is explained in details in section~\ref{sec:const}; briefly, weak spatially-confined forcings of zonal-wind and temperature are applied in the dry dynamical core with Held-Suarez physics at $100$ latitude-pressure combinations one at a time. For each forcing, the time-mean response $(\langle \overline{u} \rangle,\langle \overline{T} \rangle)$ is calculated from a long integration, and the responses and imposed forcings from all runs are then used to find the LRF from the long-time averaged Eq.~(\ref{eq:lrf0}) using matrix inversion. Similarly, $\langle \overline{u'v'} \rangle$ and $\langle \overline{v'T'} \rangle$ are calculated for all the runs and along with time-mean responses are used to find the EFM from Eq.~(\ref{eq:efm}) using matrix inversion. 

The remainder of this paper is structured as follows. In section~\ref{sec:eqs} we discuss Eqs.~(\ref{eq:lrf0})-(\ref{eq:efm}) and the underlying assumptions, followed by descriptions of the model setup and the detailed procedure for calculating the LRF and EFM in section~\ref{sec:const}. Several tests to validate the calculated LRF and EFM are presented in section~\ref{sec:test} which show that the LRF is fairly skillful in finding the time-mean response to a given zonally-symmetric external forcings and vice versa, and that the EFM can quantify the time-mean response of eddy fluxes to a change in the mean-flow reasonably well. In section~\ref{sec:nv} we present some of the properties of the LRF of the idealized dry atmosphere. In particular, we show that the neutral vector (i.e., the most excitable dynamical mode) is fairly similar to the Annular Modes, which offers some insight into the reason behind the ubiquity of Annular Mode-like patterns in the response of the real and modeled atmospheres to forcings. Potential applications of the calculated LRF and EFM are discussed in section~\ref{sec:app}, which include i) forcing a specified mean-flow for hypothesis-testing in the idealized GCM, e.g., as used in \citet{pedram15} to probe causality in the relationship between the negative phase of Arctic Oscillation and increased blocking, ii) isolating and quantifying the eddy-feedbacks in complex eddy-mean flow interaction problems, and iii) examining and evaluating more generally-applicable methods that are currently employed to diagnose eddy-feedbacks or construct LRFs in more complex GCMs, e.g., as used in Part~2 of this study to investigate why the LRF constructed using the FDT performs poorly in some cases. The paper is summarized in section~\ref{sec:sum}.        

\section{Formulation}\label{sec:eqs}
We start with the derivation of Eq.~(\ref{eq:lrf0}) for zonally-averaged forcings/responses, which, although straightforward, helps with better understanding the underlying assumptions. The zonally-averaged equations of the climate system can be written as 
\begin{eqnarray}
\dot{\overline{\mathrm{\mathbf{X}}}} = \mathds{F}(\overline{\mathrm{\mathbf{X}}})
\label{eq:gen}
\end{eqnarray}
where the state-vector $\overline{\mathrm{\mathbf{X}}}(t)$  is a set of zonally-averaged variables that we assume can uniquely describe the system and $\mathds{F}$ is a nonlinear function that represents the relevant physical processes and describes the evolution of $\overline{\mathrm{\mathbf{X}}}$. For the purpose of this paper, Eq.~(\ref{eq:gen}) is the zonally-averaged primitive equations with Held-Suarez physics \citep{held1994proposal} with the additional assumption that eddies (defined as deviations from the zonal-mean) are in statistical-equilibrium with the state-vector $\overline{\mathrm{\mathbf{X}}}$ so that the eddy fluxes can be uniquely determined from $\overline{\mathrm{\mathbf{X}}}$. This assumption and the other assumptions related to the state-vector are further discussed later in this section. 

If the system's state-vector evolves from $\overline{\mathrm{\mathbf{X}}}(0) = \langle \overline{\mathrm{\mathbf{X}}} \rangle$ to $\overline{\mathrm{\mathbf{X}}}(t) = \langle \overline{\mathrm{\mathbf{X}}} \rangle+\overline{\mathrm{\mathbf{x}}}(t)$ in response to an external forcing $\overline{\mathrm{\mathbf{f}}}(t)$ of zonal torque or buoyancy, where $\langle \overline{\mathrm{\mathbf{X}}} \rangle$ is the mean-flow of the unforced state-vector, Eq.~(\ref{eq:gen}) becomes
\begin{eqnarray}
\dot{\overline{\mathrm{\mathbf{x}}}} = {\mathds{F}} \left(\langle \overline{\mathrm{\mathbf{X}}} \rangle+\overline{\mathrm{\mathbf{x}}} \right) + \overline{\mathrm{\mathbf{f}}}
\label{eq:gen2}
\end{eqnarray}
It should be clarified that in this section and the rest of the paper ``mean-flow'' and $\langle \cdot \rangle$ denote very long-term averages so that ${\mathds{F}}(\langle \overline{\mathrm{\mathbf{X}}} \rangle) \approx 0$. A Taylor expansion of ${\mathds{F}}$ around $\langle \overline{\mathrm{\mathbf{X}}} \rangle$ yields Eq.~(\ref{eq:lrf0}) for the state-vector response $\overline{\mathrm{\mathbf{x}}}$, which as mentioned earlier is a set of zonal-mean variables:
\begin{eqnarray}
\dot{\overline{\mathrm{\mathbf{x}}}} = \left. \frac{d{\mathds{F}}}{d \overline{\mathrm{\mathbf{X}}}}\right|_{\langle\overline{\mathrm{\mathbf{X}}}\rangle} \, \overline{\mathrm{\mathbf{x}}} + \overline{\mathrm{\mathbf{f}}} = \pmb{\mathsf{L}} \, \overline{\mathrm{\mathbf{x}}} + \overline{\mathrm{\mathbf{f}}}
\label{eq:lrf1}
\end{eqnarray}
where we have assumed that the terms of order $\overline{\mathrm{\mathbf{x}}}^2$ and higher are negligible compared to the term that is linear in $\overline{\mathrm{\mathbf{x}}}$ (assumption 1). The LRF $\pmb{\mathsf{L}}$ is thus the Jacobian of ${\mathds{F}}(\overline{\mathrm{\mathbf{X}}})$ evaluated at $\langle \overline{\mathrm{\mathbf{X}}}\rangle$. See \citet{palmer99} and \citet{farrell1996generalized,farrell1996generalizedII} for further discussions of the above equations from a dynamical-system perspective. 

It should be noted that ignoring terms that are nonlinear in $\overline{\mathrm{\mathbf{x}}}$ does not eliminate the eddy-feedbacks from $\pmb{\mathsf{L}}$ in Eq.~(\ref{eq:lrf1}), and the difference between this equation and the equation that could be derived following the first approach discussed in section~\ref{sec:intro} (i.e., linearizion of the equations of motion) should be emphasized: the eddy fluxes are absent in the latter while they are linearized as a function of $\overline{\mathrm{\mathbf{x}}}$ in (\ref{eq:lrf1}). For such representation to be valid, we have to assume that the eddy statistics are in quasi-equilibrium with $\overline{\mathrm{\mathbf{x}}}$, or said another way, the synoptic eddies respond to changes in the mean-flow at timescales that are much shorter compared to the timescales of changes in the mean-flow  \citep{lorenz2001eddy,ring08}. This is justified if the variability of $\overline{\mathrm{\mathbf{x}}}$ has timescales of several days or longer (assumption 2). To satisfy this requirement, we assume that $\overline{\mathrm{\mathbf{x}}}$ is the anomaly averaged over a few days, which is appropriate for zonal-means in many large-scale phenomena of interest. 

We also note the difference between our approach to finding the LRF in Eq.~({\ref{eq:lrf1}}) and the second approach discussed in section~\ref{sec:intro}, where the eddy fluxes are parametrized (as done here) but the relationship between the eddy fluxes and the mean-flow is assumed to be known from a turbulence closure approximation. Here we calculate the LRF from a fully nonlinear eddy-resolving model instead of making such assumption about this relationship. 

It should be further clarified that the external forcing $\overline{\mathrm{\mathbf{f}}}$ refers to any mechanical or thermal forcing exerted by unresolved or unrepresented processes and phenomena, including climate change-induced forcings such as high-latitude warming (i.e., Arctic Amplification), tropical tropospheric warming (due to latent heating), and stratospheric cooling (due to ozone depletion) \citep{butler2010steady}. In light of the above assumption on the timescale of $\overline{\mathrm{\mathbf{x}}}$, external forcing $\overline{\mathrm{\mathbf{f}}}$ may also represent the stochastic eddy forcing generated by the atmospheric internal dynamics at (unresolved) short timescales and small spatial-scales.   

The next question to answer is what variables constitute $\overline{\mathrm{\mathbf{x}}}$. The state-vector ${\mathrm{\mathbf{X}}}$ of the primitive equations of the Held-Suarez setup involve four prognostic variables: zonal $U$ and meridional $V$ velocities, temperature $\Theta$, and surface pressure $P_s$ (diagnostic variables such as vertical velocity $\Omega$ and geopotential height $\Phi$ can be calculated from the continuity equation and hydrostatic balance later). Although the zonal-mean responses of these four variables $(\overline{u},\overline{v},\overline{T},\overline{p}_s)$ can be used for $\overline{\mathrm{\mathbf{x}}}$, it is desirable to reduce the number of variables to lower the computational cost of finding the LRF and EFM.

%, because the procedure used here (section~\ref{sec:const}) requires forcing each variable of $\overline{\mathrm{\mathbf{x}}}$ at $100$ latitude-pressure combinations and conducting long integrations to increase the signal-to-noise ratio. 

To reduce  $\overline{\mathrm{\mathbf{x}}}$, first we highlight that the zonally-averaged mean-flow of the response consists of $\overline{u}$ and $\overline{T}$, which are closely in gradient-wind balance except near the equator, and $\overline{v}$ and $\overline{\omega}$, which form the meridional circulation and can be represented by a single streamfunction $\chi$ via the continuity equation. As outlined in Appendix~A, we can choose 
\begin{eqnarray}
\overline{\mathrm{\mathbf{y}}} \equiv 
\begin{bmatrix}
    \overline{u} \\
    \overline{T}
  \end{bmatrix}
\label{eq:state}
\end{eqnarray}
if (a) $\overline{u}$ and $\overline{T}$ are in gradient-wind balance (which is very reasonable outside the deep tropics), and/or if (b) the tendencies of $\overline{u}$ and $\overline{T}$ in the zonal-momentum and temperature equations are negligible compared to the dominant terms (which is very reasonable given assumption 2). If (b) is satisfied the flow is also likely to be in close gradient-wind balance; however, here we discuss both (a) and (b) to emphasize the only place where the gradient-wind balance may play a role in this formulation (see Appendix~A for further discussions). With either (a) or (b), $\chi$ is in quasi-equilibrium with $\overline{\mathrm{\mathbf{y}}}$ and can be determined for a given ($\overline{u},\overline{T},\bar{f}$) from a diagnostic equation, and the system can be completely defined by $\overline{\mathrm{\mathbf{y}}} \equiv (\overline{u},\overline{T})$. As a result, the state-vector reduces to (\ref{eq:state}), and Eq.~(\ref{eq:lrf1}) reduces to
\begin{eqnarray}
\dot{\overline{\mathrm{\mathbf{y}}}} = \pmb{\mathsf{M}} \, \overline{\mathrm{\mathbf{y}}} + \overline{\mathrm{\mathbf{f}}}_\mathrm{eff} 
\label{eq:lrf3}
\end{eqnarray}
where the size and elements of $\pmb{\mathsf{M}}$, the LRF of the reduced system (\ref{eq:lrf3}), are different from $\pmb{\mathsf{L}}$, but they contain the same physics: Rayleigh drag, Newtonian cooling, eddy-feedback, meridional advection of $(\overline{u},\overline{T})$ by $(\langle \overline{V} \rangle,\langle \overline{\Omega} \rangle)$, and meridional advection of $(\langle \overline{U} \rangle,\langle \overline{\Theta} \rangle)$ by $\chi$. The effective external forcing $\overline{\mathrm{\mathbf{f}}}_\mathrm{eff}= \pmb{\mathsf{B}} \, \overline{\mathrm{\mathbf{f}}}$ has size, elements, and physics that are different from $\overline{\mathrm{\mathbf{f}}}$: the latter is the direct external forcing in $(\overline{U},\overline{\Theta})$ that is applied to the atmosphere, while the effective forcing includes this direct forcing plus changes in $\chi$ due to the component of $\overline{\mathrm{\mathbf{f}}}$ that is not in gradient-wind balance. This is related to the classic Eliassen balanced flow problem \citep{eliassen1951slow} and is further discussed in Appendix~A. We are mainly interested in finding  
\begin{eqnarray}
\tilde{\pmb{\mathsf{M}}} = \pmb{\mathsf{B}}^{-1} \, \pmb{\mathsf{M}} 
\label{eq:lrf4}
\end{eqnarray}
which relates time-mean response $\langle \overline{\mathrm{\mathbf{y}}} \rangle$ to the direct external forcing $\langle \overline{\mathrm{\mathbf{f}}} \rangle$ as                      
\begin{eqnarray}
\tilde{\pmb{\mathsf{M}}} \langle \overline{\mathrm{\mathbf{y}}}  \rangle = - \langle \overline{\mathrm{\mathbf{f}}}  \rangle. 
\label{eq:steady}
\end{eqnarray}

In summary, Eqs.~(\ref{eq:state})-(\ref{eq:steady}) require the following assumptions:
\begin{enumerate}
\item The forcing is weak enough so that the system is in the linear regime (i.e., $\overline{\mathrm{\mathbf{y}}}$ depends linearly on $\overline{\mathrm{\mathbf{f}}}$).    
\item The variability of $\overline{\mathrm{\mathbf{y}}}$ has timescales of several days or longer, so that the synoptic eddies and the meridional circulation are in quasi-equilibrium with $\overline{\mathrm{\mathbf{y}}}$.    
\end{enumerate}

Finally it is worth mentioning that potential vorticity (along with lower boundary conditions) is another appropriate state-vector \citep[see, e.g.,][]{hoskins1985use}; however, here have chosen to use $(\overline{u},\overline{T})$ for practical convenience. 

\section{Construction of the LRF and EFM} \label{sec:const}
\subsection{Idealized Dry GCM} \label{subsec:model}
We use the GFDL dry dynamical core, which is a pseudo-spectral GCM that solves the primitive equations on sigma levels $\sigma$. The model is used with the Held-Suarez setup, which is described in detail in \citet{held1994proposal}. Briefly, the model is forced by Newtonian relaxation of temperature to a prescribed equinoctial radiative-equilibrium state with a specified equator-to-pole surface temperature difference of $60$~K. The relaxation timescale is $40$~days except at the lower-level tropics where it is changed to $4$~days to create a more realistic Hadley circulation and to prevent a boundary-layer temperature inversion \citep{held1999surface}. Rayleigh drag with a prescribed rate, which decreases linearly from $1$~day$^{-1}$ at the surface ($\sigma=1$) to zero at $\sigma=0.7$ and higher levels, is used to remove momentum from the low levels, and $\nabla^8$ hyper-diffusion is used to remove enstrophy at small scales. The forcings, dissipations, and boundary conditions are all zonally-symmetric and symmetric between the two hemispheres. A T63 spectral resolution ($\sim 1.9^\mathrm{o} \times 1.9^\mathrm{o}$) with $40$ equally-spaced sigma levels and $15$~min time-steps are used to solve the equations. Unless noted otherwise, every run is $45000$~days with the last $44500$~days used to calculate time-means denoted with $\langle \cdot \rangle $. ``Ensemble" refers to three runs with identical setup and slightly different initial conditions.                 

\subsection{Procedure}
Instead of forcing the model at every pressure $p$ (40 levels) and latitude $\mu$ (96 grid points), we use a set of $100$ basis functions with coarser resolution to reduce the computational cost. The basis functions are of the Gaussian form  
\begin{eqnarray}
\exp \left[-\frac{(|\mu|-\mu_o)^2}{\mu_w^2}-\frac{(p-p_o)^2}{p_w^2} \right]
\label{eq:forcing}
\end{eqnarray}
where $\mu_w=10^\mathrm{o}$, $\mu_o=0^\mathrm{o}, 10^\mathrm{o}, 20^\mathrm{o}, \dots 90^\mathrm{o}$, $p_w = 75$~hPa, and $p_o=100, 200, 300, \dots 1000$~hPa ($p$ is the full-level pressure in the model). Zonally-symmetric time-invariant forcings of $U$ or $\Theta$ are added at each basis function one at a time. To be clear, this forcing is added to the right-hand side of the zonal-momentum or temperature equation in the GCM and is therefore the direct external forcing $\overline{\mathrm{\mathbf{f}}}$ (not the effective forcing). Each of these forced runs are referred to as a ``trial'' hereafter. Note that the forcings are added to both hemispheres simultaneously. The amplitude of forcing $\bar{f}_o$ in each trial is chosen to obtain a large signal-to-noise ratio within the linear regime (see Appendix~B for details). 
%External torque at the highest pressure level ($\sim 9$~hPa) is found to result in unrealistically large responses in the tropical stratosphere. To avoid this problem, the basis functions of zonal-wind for $p_o=100$~hPa are set to zero at the highest pressure level and the amplitude of forcing is also reduced. 
Each trial is used to calculate $\langle \overline{U} \rangle$ and $\langle \overline{\Theta} \rangle$ interpolated on $39$ pressure levels ($25, 50, 75, \dots 975$~hPa). For each variable ($U$ or $\Theta$), two trials with $\pm \bar{f}_o$ are run for each $(\mu_o,p_o)$ and the results are combined as $\langle \overline{u} \rangle = (\langle \overline{U} \rangle_{+} - \langle \overline{U} \rangle_{-})/2$ and $\langle \overline{T} \rangle = (\langle \overline{\Theta} \rangle_{+} - \langle \overline{\Theta} \rangle_{-})/2$ where the subscript $+ (-)$ denotes the trial with $+\bar{f}_o (-\bar{f}_o)$. Combining the results cancels the quadratic terms in the Taylor expansion of Eq.~(\ref{eq:gen2}) and improves the accuracy of Eq.~(\ref{eq:lrf1}). A total of $400$ trials are needed to calculate the LRF and EFM. An ensemble of unforced simulations (referred to as the control-run and denoted with subscript $c$) is run as well.  

Once the $400$ acceptable trials are chosen following the quantitative and qualitative criteria described in Appendix~B, the hemispherically-averaged results of the trials with positive and negative forcings are combined to obtain $200$ sets of $(\langle \overline{u} \rangle,\langle \overline{T} \rangle)$. The $(\langle \overline{u} \rangle,\langle \overline{T} \rangle)$ of each set are then projected onto the $100$ basis functions~(\ref{eq:forcing}) using least-square linear regression. This results in $\langle\overline{\mathrm{\mathbf{u}}}\rangle_n$ and $\langle\overline{\mathrm{\mathbf{T}}}\rangle_n$ which are column vectors of length $100$ containing the regression coefficients of the response to forcing $\overline{\mathrm{\mathbf{f}}}_n$ ($n=1,2, \dots 200$). 
%Note that because $\langle \overline{u} \rangle$ and $\langle \overline{T} \rangle$ are hemispherically averaged, the basis functions~(\ref{eq:forcing}) should be also averaged between the hemispheres for the purpose of the projection. 
$\overline{\mathrm{\mathbf{f}}}_n$ is a column vector of length $100$ whose elements are all zero except for its $n^\mathrm{th}$ element if $n \leq 100$ or $(n-100)^\mathrm{th}$ element if $n > 100$. The non-zero element of $\overline{\mathrm{\mathbf{f}}}_n$ is the forcing amplitude of the $n$th basis function (forcing for $n=1-100$ is imposed torque and for $n=101-200$ is thermal forcing). We then assemble the response matrix $\pmb{\mathsf{R}}$ (\ref{eq:R}) and the forcing matrix $\pmb{\mathsf{F}}$ (\ref{eq:F}), which are $200 \times 200$ matrices:            
\begin{eqnarray}
\!\!\!\!\!\!\!\!\!\!\!\!\!\!\!\!\!\!\!\!\!\!\!\!  \pmb{\mathsf{R}} \!\!\!\!\!\! &=& \!\!\!\!\!\! 
\begin{bmatrix}
\label{eq:R}
    \langle\overline{\mathrm{\mathbf{u}}}\rangle_{1} & \langle\overline{\mathrm{\mathbf{u}}}\rangle_2 & \cdots &  \langle\overline{\mathrm{\mathbf{u}}}\rangle_{101} & \cdots & \langle\overline{\mathrm{\mathbf{u}}}\rangle_{200} \\
   \, \langle\overline{\mathrm{\mathbf{T}}}\rangle_1 & \langle\overline{\mathrm{\mathbf{T}}}\rangle_2 & \cdots &  \langle\overline{\mathrm{\mathbf{T}}}\rangle_{101} & \cdots & \langle\overline{\mathrm{\mathbf{T}}}\rangle_{200} 
  \end{bmatrix}\\
%\end{eqnarray}
%\begin{eqnarray}
\nonumber \\ 
\!\!\!\!\!\!\!\!\!\!\!\!\!\!\!\!\!\!\!\!\!\!\!\! \pmb{\mathsf{F}} \!\!\!\!\!\! &=& \!\!\!\!\!\!  
\begin{bmatrix}
    \;\;\;\overline{\mathrm{\mathbf{f}}}_{1}  & \;\;\;\;\; \overline{\mathrm{\mathbf{f}}}_{2} & \;\;\, \cdots & \;\; {\mathrm{\mathbf{0}}}       & \;\;\; \cdots  &  \;\;\; {\mathrm{\mathbf{0}}} \;\; \\
    \;\;\;{\mathrm{\mathbf{0}}}      & \;\;\;\;\; {\mathrm{\mathbf{0}}}     & \;\;\, \cdots & \;\;  \overline{\mathrm{\mathbf{f}}}_{101} & \;\;\; \cdots  & \;\;\;  \overline{\mathrm{\mathbf{f}}}_{200} \;\;
  \end{bmatrix}
\label{eq:F}
\end{eqnarray}
where $\mathrm{\mathbf{0}}$ is a $100 \times 1$ zero matrix, and $\pmb{\mathsf{F}}$ is a diagonal matrix. Equation~(\ref{eq:steady}) is then used to calculate $\tilde{\pmb{\mathsf{M}}}$: 
\begin{eqnarray}
\tilde{\pmb{\mathsf{M}}} =  -\pmb{\mathsf{F}} \, \pmb{\mathsf{R}}^{-1}.
\label{eq:Mgrf} 
\end{eqnarray}  

As discussed in \citet{kuang2010linear}, because (\ref{eq:Mgrf}) involves $\pmb{\mathsf{R}}^{-1}$, the eigenvalues of $\tilde{\pmb{\mathsf{M}}}$ (denoted as $\lambda$) have uncertainties ($\delta \lambda$) that scale as $|\delta \lambda| \propto \lambda^2 \| \delta \pmb{\mathsf{R}} \|$, where $\| \cdot \|$ is a matrix norm and $\delta \pmb{\mathsf{R}}$ is the errors in $\pmb{\mathsf{R}}$ (note that $\pmb{\mathsf{F}}$ is imposed and hence precisely known). As a result, the errors in $\pmb{\mathsf{R}}$ have the least (most) influence on the eigenvalues of $\tilde{\pmb{\mathsf{M}}}$ with the smallest (largest) magnitude, which are calculated with the highest (lowest) accuracy. In fact, recalculating $\tilde{\pmb{\mathsf{M}}}$ using several of the trials replaced with runs with slightly different $\bar{f}_o$ results in substantial changes in the modes with large $|\lambda|$, while the modes with small $|\lambda|$ are robust. It is a desired property that the slowest decaying modes (with timescales of a few days and longer) are accurately calculated, because these are the modes relevant to the large-scale circulation (e.g., Annular Modes have timescales on the order of tens of days). Fast modes (timescale of $1$~day or shorter), which are inaccurately calculated and also violate assumption 2 (i.e., quasi-equilibrium), have timescales much shorter than that of the large-scale circulation. All but five of the eigenmodes of the computed $\tilde{\pmb{\mathsf{M}}}$ are decaying ($\lambda$ has negative real part) where the slowest decaying mode has $\lambda \sim -0.017$~day$^{-1}$. The eigenvalues of $\tilde{\pmb{\mathsf{M}}}$ with timescales longer than $1$~day are shown in Fig.~S1 (Supplemental Material) and the eigenvalues/vectors are further discussed in section~\ref{sec:nv}. The five growing eigenmodes have $\lambda$ with real and imaginary parts on the orders of $100-1000$~day$^{-1}$, which is much faster than the shortest timescale resolved in the GCM (i.e., the $15$~min timestep).         

The five growing modes are evidently erroneous and following \citet{kuang2010linear}, the sign of eigenvalues for these modes are reversed and the matrix is reconstructed using the eigenvectors/eigenvalues. Hereafter, $\tilde{\pmb{\mathsf{M}}}$ refers to this matrix, which only has decaying modes. The rapidly decaying modes with large $|\lambda|$ on the order of $10-1000$~day$^{-1}$ are also inaccurate. These modes do not affect calculations that involve $\tilde{\pmb{\mathsf{M}}}^{-1}$ multiplication (such as calculating the response to a given forcing); however, they degrade the accuracy of calculations that involve $\tilde{\pmb{\mathsf{M}}}$  multiplication. For example, these modes can result in unphysically large forcings calculated for a given response. This problem is solved by filtering out the fast-decaying inaccurate modes by calculating matrix ${\hat{\pmb{\mathsf{M}}}}$ as                          
\begin{eqnarray}
\hat{\pmb{\mathsf{M}}} = \frac{\exp\left[\tilde{\pmb{\mathsf{M}}} \epsilon \right]-\pmb{\mathsf{I}}}{\epsilon}
\label{eq:M2} 
\end{eqnarray}
where $\epsilon=1$~day and $\pmb{\mathsf{I}}$ is the identity matrix. This procedure filters modes with timescale faster than $\epsilon$ while leaving modes with slower timescales almost intact. $\hat{\pmb{\mathsf{M}}}$ can be interpreted as $\tilde{\pmb{\mathsf{M}}}$ averaged over time $\epsilon$.          

To find $\pmb{\mathsf{E}}$, we first compute the eddy fluxes for each trial using the anomalous (with respect to the climatology of each trial) daily-averaged zonal and meridional winds and temperature at a lower resolution (every other grid-point in latitude, longitude, and pressure) for computational tractability. Results from trials with positive and negative forcings are combined to find $\langle\overline{{u'v'}}\rangle$ and $\langle\overline{{v'T'}}\rangle$, which are then projected onto the basis functions. Matrix $\pmb{\mathsf{Q}}$ is calculated following the same procedure used for $\pmb{\mathsf{R}}$:    
\begin{eqnarray}
\!\!\!\!\!\!\!\!\!\!\!\! \pmb{\mathsf{Q}} \!\!\!\!\!\! &=& \!\!\!\!\!\! 
\begin{bmatrix}
    \langle\overline{\mathrm{\mathbf{u'v'}}}\rangle_{1} \!\!\! & \!\!\! \langle\overline{\mathrm{\mathbf{u'v'}}}\rangle_2 \!\!\! & \!\!\!  \cdots \!\!\! & \!\!\!  \langle\overline{\mathrm{\mathbf{u'v'}}}\rangle_{101} \!\!\! & \!\!\! \cdots \!\!\! & \!\!\! \langle\overline{\mathrm{\mathbf{u'v'}}}\rangle_{200} \\
    \langle\overline{\mathrm{\mathbf{v'T'}}}\rangle_1 \!\! & \!\! \langle\overline{\mathrm{\mathbf{v'T'}}}\rangle_2 \!\! & \!\!\! \cdots \!\!\! & \!\!\!  \langle\overline{\mathrm{\mathbf{v'T'}}}\rangle_{101} \!\!\! & \!\!\! \cdots \!\!\! & \!\!\! \langle\overline{\mathrm{\mathbf{v'T'}}}\rangle_{200} 
  \end{bmatrix} \;\;
\label{eq:Q}
\end{eqnarray}
We do not calculate $\pmb{\mathsf{E}}$ using $\pmb{\mathsf{Q}} \, \pmb{\mathsf{R}}^{-1}$, because its large eigenvalues will be inaccurate (as discussed above) and the appropriate threshold $\epsilon$ for filtering similar to Eq.~(\ref{eq:M2}) is unclear. Instead, we use $\hat{\pmb{\mathsf{M}}}$, which is already filtered, as  
\begin{eqnarray}
\hat{\pmb{\mathsf{E}}} =  - \pmb{\mathsf{Q}} \, \left( \pmb{\mathsf{F}}^{-1} \, \hat{\pmb{\mathsf{M}}} \right)
\label{eq:Egrf} 
\end{eqnarray}       
which can be interepreted as the EFM averaged over time $\epsilon$.  

We have not systematically attempted to optimize various aspects of the procedure such as the basis functions (\ref{eq:forcing}), forcing amplitudes $\bar{f}_o$ (Tables~S1-S2), or the filtering step (\ref{eq:M2}). Still, as shown using several tests in the next section, the calculated LRF and EFM are fairly accurate and skillful.    

\section{Validation of the LRF and EFM} \label{sec:test}
In this section we use three tests with varying degrees of complexity to examine the performance of $\hat{\pmb{\mathsf{M}}}$ in calculating the time-mean response to an external forcing or vice versa and compare the results with those produced using the GCM. We use the same test cases to investigate the accuracy of $\hat{\pmb{\mathsf{E}}}$ in computing the changes in eddy momentum and heat fluxes in response to a given change in the mean-flow (subsection~\ref{sec:test}.\ref{sec:testE}). 
%The three tests are 1) mean-flow response to an imposed Gaussian thermal forcing; 2) forcing needed for a mean-flow response that matches the change in the mean-flow when the Newtonian relaxation time is increased by $10 \%$; 3) forcing needed for a mean-flow response that matches the leading pattern of the internal variability of the model (i.e., the Annular Mode).

\subsection{Test 1: Mean-flow response to an imposed Gaussian thermal forcing}\label{sec:test1}
\begin{figure*}[t]
\centerline{\includegraphics[width=1\textwidth]{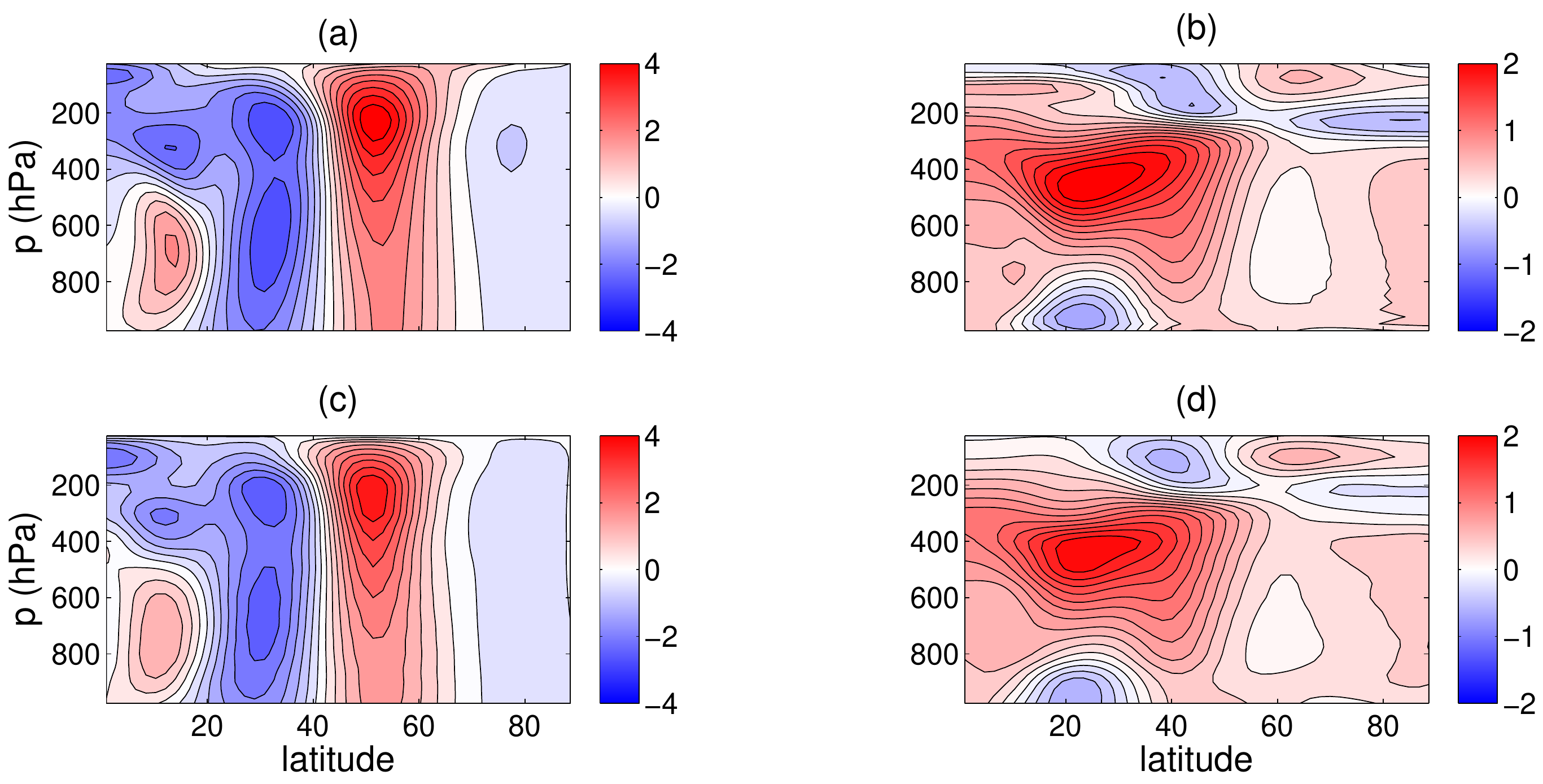}}
\caption{Results from Test~1: time-mean response to an imposed Gaussian thermal forcing $\bar{f}$ (see subsection~\ref{sec:test}.\ref{sec:test1} for details). The ensemble-mean forced response, with respect to the ensemble-mean control-run, calculated from the GCM forced with $\bar{f}$ is shown in (a) $\langle \overline{u}\rangle_\mathrm{GCM}$ in m~s$^{-1}$ and (b) $\langle \overline{T}\rangle_\mathrm{GCM}$ in K. (c) $\langle \overline{u}\rangle_\mathrm{LRF}$ in m~s$^{-1}$ and (d) $\langle \overline{T}\rangle_\mathrm{LRF}$ in K show the response to $\bar{f}$ calculated using the LRF, $\hat{\pmb{\mathsf{M}}}$. Employing the norms defined in Eqs.~(\ref{eq:max}) and (\ref{eq:l2}), relative errors are $| \|\langle \overline{u} \rangle_\mathrm{LRF}   \|_\infty - \| \langle \overline{u} \rangle_\mathrm{GCM}\|_\infty | \times 100 / \| \langle \overline{u}\rangle_\mathrm{GCM}\|_\infty = 14  \%$, 
$| \|\langle \overline{T} \rangle_\mathrm{LRF}   \|_\infty - \| \langle \overline{T} \rangle_\mathrm{GCM}\|_\infty | \times 100 / \| \langle \overline{T}\rangle_\mathrm{GCM}\|_\infty = 8  \%$,
$  \|\langle \overline{u} \rangle_\mathrm{LRF}             -    \langle \overline{u} \rangle_\mathrm{GCM}\|_2        \times 100 / \| \langle \overline{u}\rangle_\mathrm{GCM}\|_2 = 21\% $, and   
$  \|\langle \overline{T} \rangle_\mathrm{LRF}             -    \langle \overline{T} \rangle_\mathrm{GCM}\|_2        \times 100 / \| \langle \overline{T}\rangle_\mathrm{GCM}\|_2 = 15\% $.}
\label{fig:test1}
\end{figure*}

First, we test the accuracy of $\hat{\pmb{\mathsf{M}}}$ in calculating the time-mean response $\langle \overline{\mathrm{\mathbf{y}}}  \rangle$ to a simple external thermal forcing $\bar{f} =  0.2 \times \exp{\left[-(p-450)^2/125^2-(|\mu|-25)^2/15^2\right]}$ with units of K~day$^{-1}$. An ensemble of simulations forced with $\bar{f}$ is run and the ensemble-mean response is shown in Figs.~\ref{fig:test1}(a)-\ref{fig:test1}(b). To calculate the response from the LRF, $\bar{f}$ is first linearly regressed onto the $100$ basis functions (\ref{eq:forcing}) to find a column of regression coefficients $\overline{\mathrm{\mathbf{f}}}$ (note that the forcing is chosen so that it is not representable by a single basis function). The response is then calculated from $-\hat{\pmb{\mathsf{M}}}^{-1} \, \overline{\mathrm{\mathbf{f}}}$ as coefficients of the basis functions. The response in the grid-space is shown in Figs.~\ref{fig:test1}(c)-\ref{fig:test1}(d). Results show that $\hat{\pmb{\mathsf{M}}}$ is fairly accurate in calculating the amplitude of the response and its pattern, even at relatively small-scales. We have found $\hat{\pmb{\mathsf{M}}}$ similarly skillful in several other tests with thermal or mechanical Gaussian forcings.          
    
\subsection{Test 2: Forcing needed for a given mean-flow response}
\begin{figure*}[t!]
\centerline{\includegraphics[width=1\textwidth]{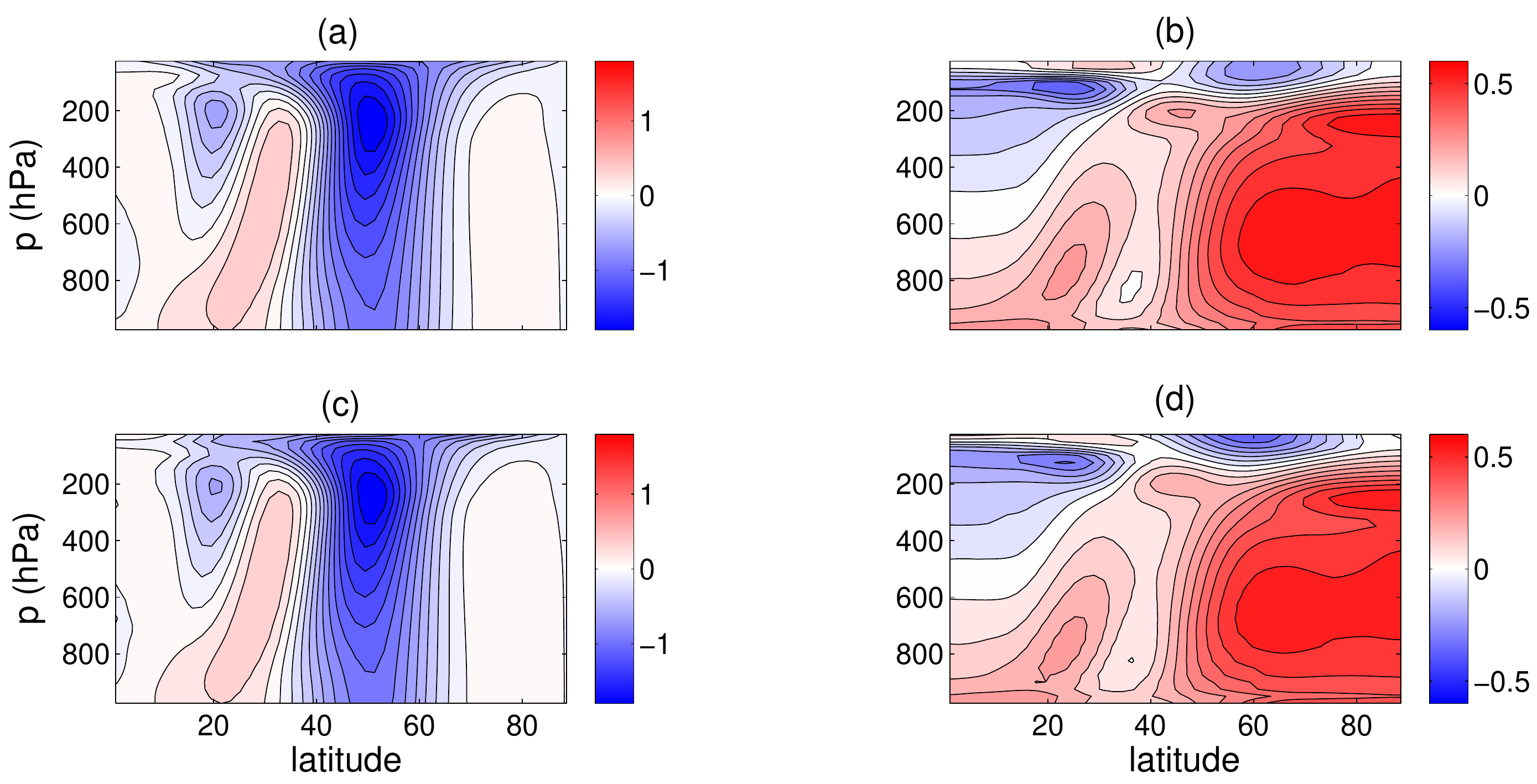}}
\caption{Results from Test~2: Forcing a specified time-mean response (target). (a) $\langle \overline{u}\rangle_\mathrm{GCM}$ in m~s$^{-1}$ and (b) $\langle \overline{T} \rangle_\mathrm{GCM}$ in K are the time-mean response of an ensemble GCM run with Newtonian relaxation timescale that is $10\%$ larger than that of the control-run. (a)-(b) are the target. (c) $\langle \overline{u}\rangle_\mathrm{LRF}$~m~s$^{-1}$ and (d) $\langle \overline{T}\rangle_\mathrm{LRF}$~K are the ensemble-mean response calculated from the GCM forced with $\overline{\mathrm{\mathbf{f}}}= - \hat{\pmb{\mathsf{M}}} \times (\langle\overline{u}\rangle_\mathrm{GCM},\langle\overline{T}\rangle_\mathrm{GCM})$ to match (a)-(b). Employing the norms defined in Eqs.~(\ref{eq:max}) and (\ref{eq:l2}), relative errors are
$| \|\langle \overline{u} \rangle_\mathrm{LRF}   \|_\infty - \| \langle \overline{u} \rangle_\mathrm{GCM}\|_\infty | \times 100 / \| \langle \overline{u}\rangle_\mathrm{GCM}\|_\infty = 1  \%$,
$| \|\langle \overline{T} \rangle_\mathrm{LRF}   \|_\infty - \| \langle \overline{T} \rangle_\mathrm{GCM}\|_\infty | \times 100 / \| \langle \overline{T}\rangle_\mathrm{GCM}\|_\infty = 3  \%$,
$  \|\langle \overline{u} \rangle_\mathrm{LRF}             -    \langle \overline{u} \rangle_\mathrm{GCM}\|_2        \times 100 / \| \langle \overline{u}\rangle_\mathrm{GCM}\|_2 = 6 \% $, and   
$  \|\langle \overline{T} \rangle_\mathrm{LRF}             -    \langle \overline{T} \rangle_\mathrm{GCM}\|_2        \times 100 / \| \langle \overline{T}\rangle_\mathrm{GCM}\|_2 = 8 \% $.}
\label{fig:test2}
\end{figure*}

In Tests 2-3, we test whether $\hat{\pmb{\mathsf{M}}}$ can accurately calculate the time-invariant forcing $\overline{\mathrm{\mathbf{f}}}$ needed to achieve a complex specified time-mean response. In Test 2, the target is the time-mean response to a $10\%$ increase in the  Newtonian relaxation timescale of the Held-Suarez setup. We run an ensemble using the GCM with the relaxation timescale increased to $44$~days and compute $\langle \overline{\mathrm{\mathbf{y}}} \rangle=(\langle \overline{u} \rangle,\langle \overline{T} \rangle)$, which is the target and shown in Figs.~\ref{fig:test2}(a)-\ref{fig:test2}(b). Time-invariant forcing needed to generate $\langle \overline{\mathrm{\mathbf{y}}} \rangle$ in a setup with the original relaxation time of $40$~days is then calculated as $\overline{\mathrm{\mathbf{f}}}= - \hat{\pmb{\mathsf{M}}} \, \langle \overline{\mathrm{\mathbf{y}}} \rangle$ and applied in the GCM (with the original setup) to run an ensemble ($\overline{\mathrm{\mathbf{f}}}$ is shown in Fig.~S2 and discussed in section~S3). The ensemble-mean response is shown in Figs.~\ref{fig:test2}(c)-\ref{fig:test2}(d) and agrees well, in amplitude and pattern, with the target.             
 
\subsection{Test 3: Forcing needed to generate the Annular Mode as the mean-flow response}
\begin{figure*}[t!]
\centerline{\includegraphics[width=1\textwidth]{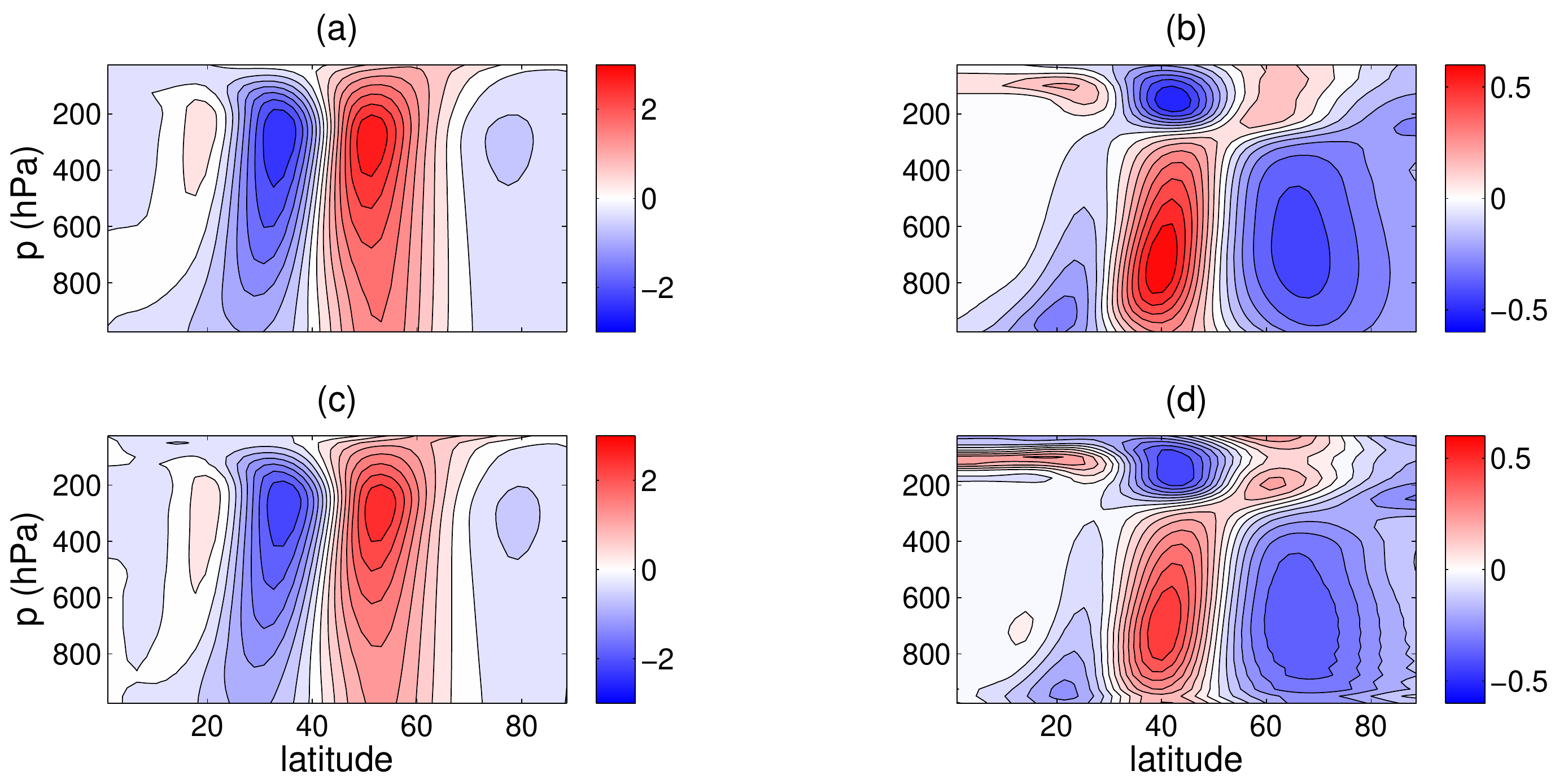}}
\caption{Results from Test~3: forcing the positive phase of Annular Mode as the time-mean response. The first EOF (EOF1) of the control-run (a) $\overline{u}_\mathrm{EOF1}$ in m~s$^{-1}$ and (b) $\overline{T}_\mathrm{EOF1}$ in K is the target. The ensemble-mean forced response (with respect to the ensemble-mean control-run) (c) $\langle \overline{u}\rangle_\mathrm{LRF}$ in m~s$^{-1}$ and (d) $\langle \overline{T}\rangle_\mathrm{LRF}$ in K are calculated from the GCM forced with $\overline{\mathrm{\mathbf{f}}}= - \hat{\pmb{\mathsf{M}}} \times \mathrm{EOF1}$ to match (a)-(b). Employing the norms defined in Eqs.~(\ref{eq:max}) and (\ref{eq:l2}), relative errors are 
$| \|\langle \overline{u} \rangle_\mathrm{LRF}   \|_\infty - \| \langle \overline{u} \rangle_\mathrm{EOF1}\|_\infty | \times 100 / \| \langle \overline{u}\rangle_\mathrm{EOF1}\|_\infty = 8  \%$,
$| \|\langle \overline{T} \rangle_\mathrm{LRF}   \|_\infty - \| \langle \overline{T} \rangle_\mathrm{EOF1}\|_\infty | \times 100 / \| \langle \overline{T}\rangle_\mathrm{EOF1}\|_\infty = 21  \%$,
$  \|\langle \overline{u} \rangle_\mathrm{LRF}             -    \langle \overline{u} \rangle_\mathrm{EOF1}\|_2        \times 100 / \| \langle \overline{u}\rangle_\mathrm{EOF1}\|_2 = 17\% $, and    
$  \|\langle \overline{T} \rangle_\mathrm{LRF}             -    \langle \overline{T} \rangle_\mathrm{EOF1}\|_2        \times 100 / \| \langle \overline{T}\rangle_\mathrm{EOF1}\|_2 = 23\% $.}
\label{fig:test3}
\end{figure*}

In Test 3, the targeted time-mean response is the positive phase of the Annular Mode of the control-run, which is calculated as the leading Empirical Orthogonal Function (EOF) of daily-averaged zonally-averaged anomalous (with respect to the climatology) zonal-wind and temperature (stacked together). The first EOF (EOF1) explains $39\%$ of the variance and is shown in Figs.~\ref{fig:test3}(a)-\ref{fig:test3}(b). Using EOF1, scaled to have $\| \langle \overline{u} \rangle_\mathrm{EOF1} \|_\infty=3$~m~s$^{-1}$, the time-invariant forcing is calculated as $\overline{\mathrm{\mathbf{f}}}= - \hat{\pmb{\mathsf{M}}} \times \mathrm{EOF1}$ and applied in the GCM to run an ensemble ($\overline{\mathrm{\mathbf{f}}}$ is shown in Fig.~S3 and discussed in section~S3). The ensemble-mean response is shown in Figs.~\ref{fig:test3}(c)-\ref{fig:test3}(d) and agrees well with the target.                           

Tests 1-3 show that the LRF $\hat{\pmb{\mathsf{M}}}$ is fairly skillful in calculating the pattern and amplitude of time-mean responses to external forcings and vice versa. These skills demonstrate that $\hat{\pmb{\mathsf{M}}}$ accurately accounts for the processes involved in the full GCM simulations and in particular the eddy-feedbacks, without which the pattern and amplitude of the forcing or response cannot be correctly captured; see section~$5$ and Fig.~15 of \citet{ring2007forced} for an example. Also it should be noted that the mean-flow changes in Tests 1-3 are not negligible fractions of the mean-flow; e.g., the amplitude of the zonal-wind change in Tests 1 and 3 is $\sim 10\%$ of the maximum climatological zonal-wind ($\sim 30$~m~s$^{-1}$), which shows that the linear approach applies to sizable forcings and responses.                

%\subsection{Test 4}
%In Test 4 the seek to calculate the forcing needed to generate a midlatitude barotropic response in the zonal-wind which is shown in Fig.~\ref{fig:test4}(a). The pattern is tapered at the very top to avoid adding torque to the top edge of the stratosphere and is nearly in gradient-wind balance without any temperature field. As before, forcing to create this time-mean response is computed using $\hat{\pmb{\mathsf{M}}}$ and applied in the GCM to run an ensemble. The ensemble-mean forced response is shown in Fig.~\ref{fig:test4}(b).     

%\begin{figure*}[th]
%\centerline{\includegraphics[width=0.85\textwidth]{Fig5hr.pdf}}
%\caption{Results from Test~4: forcing a barotropic response in zonal-wind. The target is the pattern shown in (a) $\overline{u}_\mathrm{BRT}$~m~s$^{-1}$. The time-mean response $\overline{u}_\mathrm{LRF}$ from an ensemble run with forcing calculated using $\hat{\pmb{\mathsf{M}}}$ is shown in (b). Employing the norms defined in Eqs.~\ref{eq:max} and \ref{eq:l2}, relative errors are
%$| \|\langle \overline{u} \rangle_\mathrm{LRF}   \|_\infty - \| \langle \overline{u} \rangle_\mathrm{GCM}\|_\infty | \times 100 / \| \langle \overline{u}\rangle_\mathrm{GCM}\|_\infty = 10  \%$ and
%$  \|\langle \overline{u} \rangle_\mathrm{LRF}             -    \langle \overline{u} \rangle_\mathrm{GCM}\|_2        \times 100 / \| \langle \overline{u}\rangle_\mathrm{GCM}\|_2 = 22\% $}   
%\label{fig:test4}
%\end{figure*}

\subsection{Validation of $\hat{\pmb{\mathsf{E}}}$} \label{sec:testE}
\begin{figure*}[t]
\centerline{\includegraphics[width=1\textwidth]{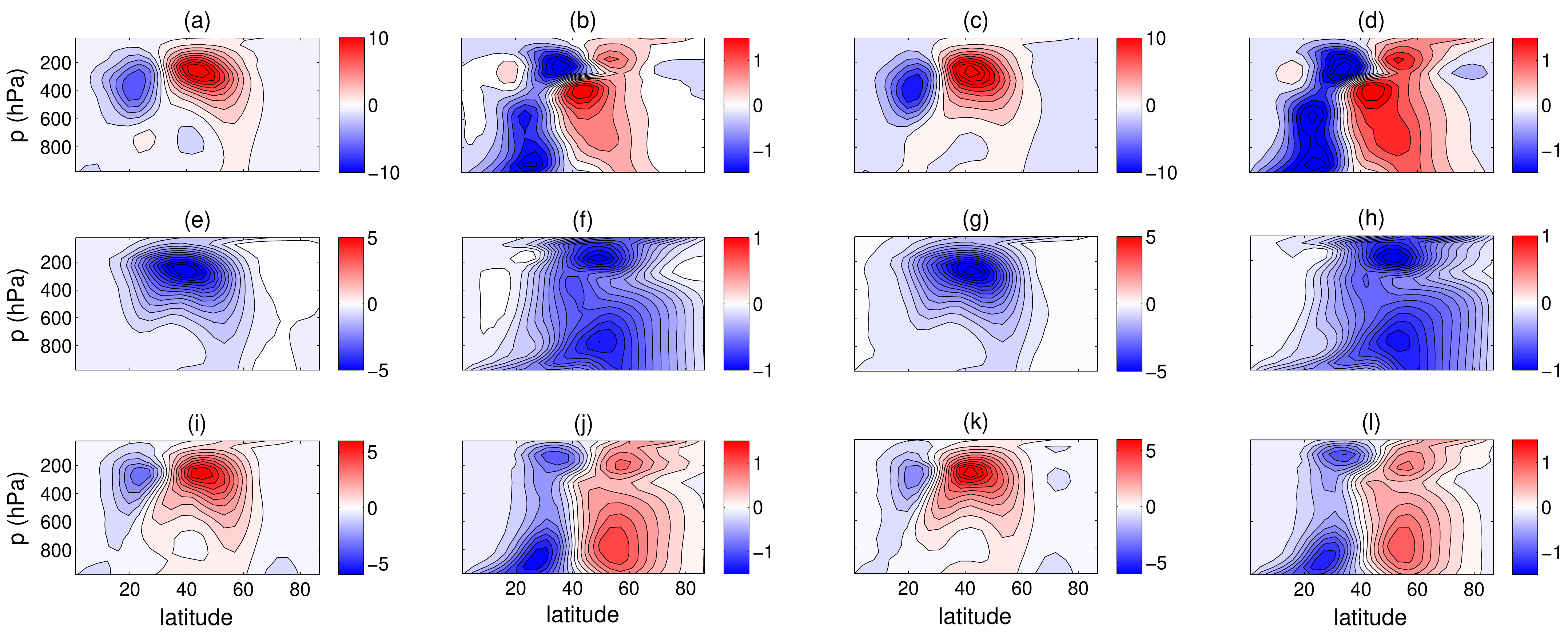}}
\caption{Changes in time-mean eddy fluxes calculated using the GCM (two left columns) and EFM (two right columns) for Tests 1-3. From the left, the first and third (second and fourth) columns show eddy momentum (heat) fluxes $\langle \overline{u'v'} \rangle$ ($\langle \overline{v'T'} \rangle$) with units m$^2$~s$^{-2}$ (K~m~s$^{-1}$). Test 1: (a)-(b) show the ensemble-mean eddy fluxes response to the forcing calculated from the GCM. (c)-(d) show the eddy fluxes response calculated using $\hat{\pmb{\mathsf{E}}}$ and the ensemble-mean $(\langle \overline{u} \rangle,\langle \overline{T} \rangle)$ of the forced GCM runs (shown in Figs.~\ref{fig:test1}(a)-\ref{fig:test1}(b)). Relative errors in amplitude are
$| \|\langle \overline{u'v'} \rangle_\mathrm{EFM}   \|_\infty - \| \langle \overline{u'v'} \rangle_\mathrm{GCM}\|_\infty | \times 100 / \| \langle \overline{u'v'}\rangle_\mathrm{GCM}\|_\infty = 27  \%$ and
$  \|\langle \overline{v'T'} \rangle_\mathrm{EFM}             -    \langle \overline{v'T'} \rangle_\mathrm{GCM}\|_\infty        \times 100 / \| \langle \overline{v'T'}\rangle_\mathrm{GCM}\|_\infty = 25\% $. Test 2: (e)-(f) show the ensemble-mean eddy fluxes response calculated from the GCM simulations with increased Newtonian relaxation time. (g)-(h) show the eddy fluxes response calculated using $\hat{\pmb{\mathsf{E}}}$ and the ensemble-mean $(\langle \overline{u} \rangle,\langle \overline{T} \rangle)$ of these GCM simulations (shown in Figs.~\ref{fig:test2}(a)-\ref{fig:test2}(b)). Relative errors in amplitude are $4\%$ and $3\%$ for the momentum and heat fluxes, respectively. Test 3: (i)-(j) show the ensemble-mean eddy fluxes response calculated from the forced GCM runs. (k)-(l) show the eddy fluxes response calculated using $\hat{\pmb{\mathsf{E}}}$ and the ensemble-mean $(\langle \overline{u} \rangle,\langle \overline{T} \rangle)$ of these forced GCM simulations (shown in Figs.~\ref{fig:test3}(c)-\ref{fig:test3}(d)). Relative errors in amplitude are $13\%$ and $11\%$ for the momentum and heat fluxes, respectively.}
\label{fig:eddy}
\end{figure*}

We use Tests 1-3 to examine the accuracy of $\hat{\pmb{\mathsf{E}}}$. In Fig.~\ref{fig:eddy}, we compare eddy fluxes $(\langle \overline{u'v'} \rangle,\langle \overline{v'T'} \rangle)$ calculated from the forced GCM simulations of Tests 1-3 with those computed using Eq.~(\ref{eq:efm}) for $(\langle \overline{u} \rangle,\langle \overline{T} \rangle)$ of these simulations (see the caption for details). These results show that the EFM, $\hat{\pmb{\mathsf{E}}}$, is skillful in calculating the amplitude and pattern of the eddy flux time-mean response to a given change in the mean-flow.         

\section{Singular Value and Eigenvalue Decompositions} \label{sec:nv}
Results of section~\ref{sec:test} validate ${\hat{\pmb{\mathsf{M}}}}$ as the accurate LRF of the idealized dry atmosphere. An important piece of information that can be obtained from ${\hat{\pmb{\mathsf{M}}}}$ is the dynamical mode with the largest response to imposed forcings. The problem can be formulated as finding the maximum of $\{ \langle \overline{\mathrm{\mathbf{y}}} \rangle,\langle \overline{\mathrm{\mathbf{y}}} \rangle \}/\{\langle \overline{\mathrm{\mathbf{f}}} \rangle,\langle \overline{\mathrm{\mathbf{f}}} \rangle \}$ where $\{ \cdot \}$ is the inner product. It follows from Eq.~(\ref{eq:steady}) that      
\begin{eqnarray}
\!\!\!\!\!\!\!\! \!\!\!\!\!\!\!\! \frac{\{ \langle \overline{\mathrm{\mathbf{y}}} \rangle,\langle \overline{\mathrm{\mathbf{y}}} \rangle \}}{\{\langle \overline{\mathrm{\mathbf{f}}} \rangle,\langle \overline{\mathrm{\mathbf{f}}} \rangle \}} \!\!\!\!\!\!\!\! &=& \!\!\!\!\!\!\!\! \frac{\{\langle \overline{\mathrm{\mathbf{y}}} \rangle,\langle \overline{\mathrm{\mathbf{y}}} \rangle \}}{\{ \tilde{\pmb{\mathsf{M}}} \, \langle \overline{\mathrm{\mathbf{y}}} \rangle, \tilde{\pmb{\mathsf{M}}} \langle \overline{\mathrm{\mathbf{y}}} \rangle \}} \!=\! \frac{\{\langle \overline{\mathrm{\mathbf{y}}} \rangle,\langle \overline{\mathrm{\mathbf{y}}} \rangle \}}{\{ \tilde{\pmb{\mathsf{M}}}^\dag \tilde{\pmb{\mathsf{M}}} \, \langle \overline{\mathrm{\mathbf{y}}} \rangle, \langle \overline{\mathrm{\mathbf{y}}} \rangle \}} \nonumber \\
 \!\!\!\!\!\!\!\! &=& \!\!\!\!\!\!\!\! \frac{\{\langle \overline{\mathrm{\mathbf{y}}} \rangle_m,\langle \overline{\mathrm{\mathbf{y}}} \rangle_m \}}{\{ \tilde{\pmb{\mathsf{M}}}^\dag \tilde{\pmb{\mathsf{M}}} \, \langle \overline{\mathrm{\mathbf{y}}} \rangle_m, \langle \overline{\mathrm{\mathbf{y}}} \rangle_m \}} \!=\! \frac{\{\langle \overline{\mathrm{\mathbf{y}}} \rangle_m,\langle \overline{\mathrm{\mathbf{y}}} \rangle_m \}}{\{ s^2_m \langle \overline{\mathrm{\mathbf{y}}} \rangle_m, \langle \overline{\mathrm{\mathbf{y}}} \rangle_m \}} 
\label{eq:nv}
\end{eqnarray}
where $\dag$ denotes the adjoint, and $\langle \overline{\mathrm{\mathbf{y}}} \rangle_m$ is the $m^\mathrm{th}$ eigenvector of $\tilde{\pmb{\mathsf{M}}}^\dag \tilde{\pmb{\mathsf{M}}}$ with eigenvalue $s^2_m$. It is evident from Eq.~(\ref{eq:nv}) that the maximum response is the eigenvector with the smallest eigenvalue, which is in fact the right singular vector of $\tilde{\pmb{\mathsf{M}}}$ with smallest singular number $s_m$, and is sometimes referred to as the neutral vector of the system  \citep{marshall1993toward,goodman2002using}. The dynamical significance of the neutral vector is that it is the largest response to forcings imposed on the atmosphere, and is therefore expected to be a prevailing component of the response to climate change-induced forcings as well as of the pattern of the low-frequency internal variability, because as discussed in section~\ref{sec:eqs}, $\overline{\mathrm{\mathbf{f}}}$ can represent the stochastic eddy forcing due to the internal atmospheric dynamics as well. See \citet{palmer99}, \citet{palmer2013singular}, \citet{farrell1996generalized,farrell1996generalizedII}, and \citet{goodman2002using} for further discussions on the significance of the singular vectors of the LRF.  

\begin{figure*}[t]
\centerline{\includegraphics[width=1\textwidth]{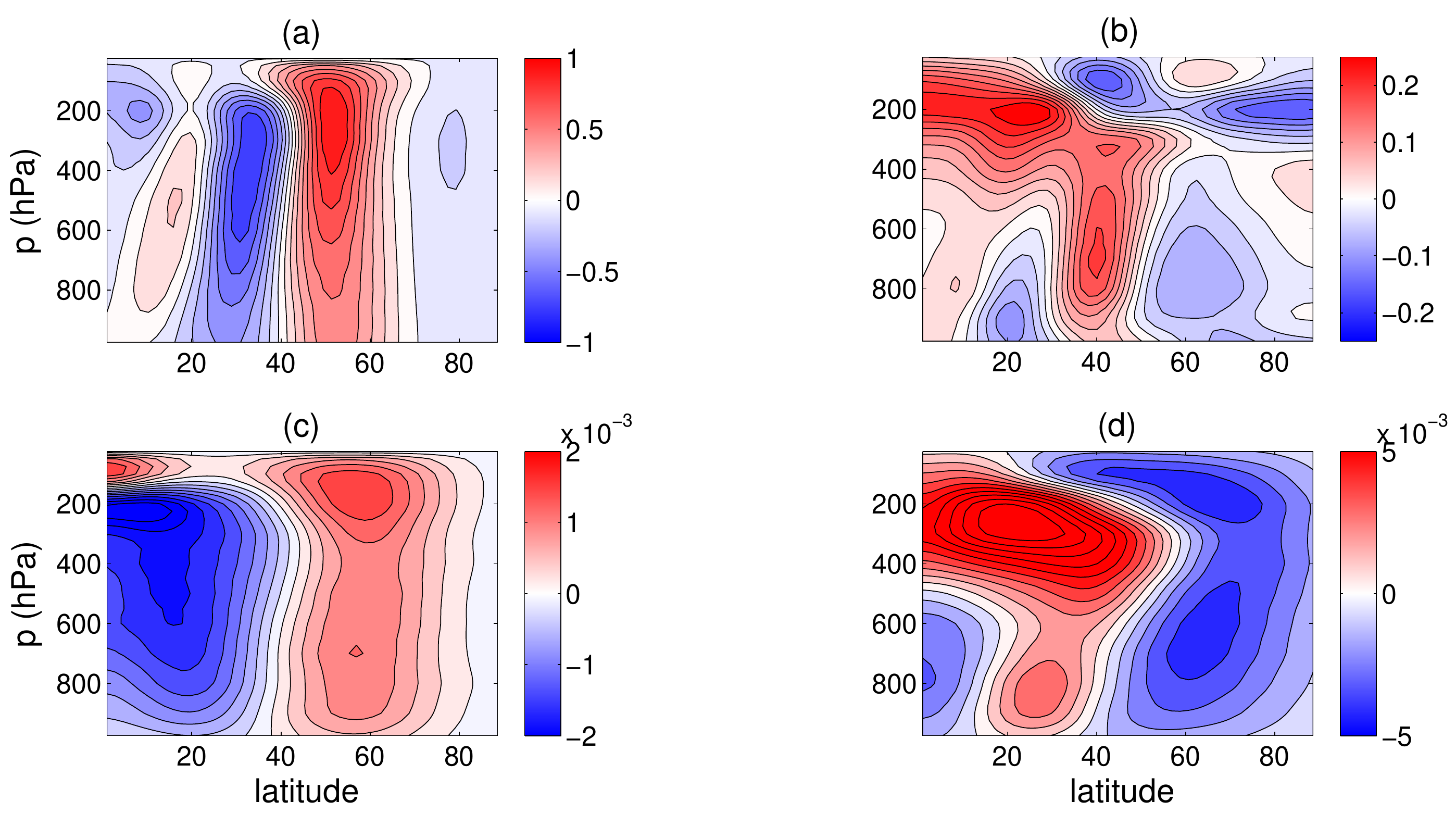}}
\caption{The neutral vector of ${\hat{\pmb{\mathsf{M}}}}$: (a) zonal-wind in m~s$^{-1}$ and (b) temperature in K. The neutral vector is calculated as the right singular vector of ${\hat{\pmb{\mathsf{M}}}}$ with the smallest singular number, and is rescaled to have $\| \overline{u} \|_\infty=1$~m~s$^{-1}$. The neutral vector of ${\tilde{\pmb{\mathsf{M}}}}$ is indistinguishable from the pattern shown here. The forcing needed to produce the neutral vector: (c) mechanical component (i.e., forcing of zonal-wind) in m~s$^{-1}$~day$^{-1}$ and (d) thermal component (i.e., forcing of temperature) in K~day$^{-1}$. The forcing is calculated as the left singular vector of ${\hat{\pmb{\mathsf{M}}}}$ with the smallest singular number (and multiplied by the singular number and rescaled in accordance with the neutral vector  rescaling). This forcing is identical to $\overline{\mathrm{\mathbf{f}}}= - \tilde{\pmb{\mathsf{M}}} \, \langle \overline{\mathrm{\mathbf{y}}} \rangle$ where $\langle \overline{\mathrm{\mathbf{y}}} \rangle$ is the rescaled neutral vector.}
\label{fig:nv}
\end{figure*}

The neutral vector of $\hat{\pmb{\mathsf{M}}}$ (and of $\tilde{\pmb{\mathsf{M}}}$) and the forcing needed to produce this pattern, both calculated using a singular value decomposition, are shown in Fig.~\ref{fig:nv}. The zonal-wind of the neutral vector is dipolar and equivalent-barotropic and strongly resembles the zonal-wind component of EOF1 (Fig.~\ref{fig:test3}(a)). This explains the dominance of Annular Mode-like patterns in the response of zonal-wind in this model to various mechanical and thermal forcings \cite[e.g.,][also see Fig.~\ref{fig:1}]{ring2007forced,butler2010steady}. These results further support the findings of \citet{ring08} that the Annular Mode is truly a dynamical mode of the system rather than just a variability pattern obtained through statistical analysis. The connection between the neutral vector and EOF1 has been discussed previously in \citet{navarra1993new}, \citet{goodman2002using}, and \citet{kuang2004norm}. In particular, as shown in \citet[][Eqs.~10-12]{goodman2002using}, the neutral vector and EOF1 are identical if the stochastic eddy forcing is spatially uncorrelated and has uniform variance everywhere, i.e., $\langle \overline{\mathrm{\mathbf{f}}} \, \overline{\mathrm{\mathbf{f}}}^\dag \rangle = \pmb{\mathsf{I}}$ if $\overline{\mathrm{\mathbf{f}}}$ has unit amplitude. The zonal-wind pattern of the neutral vector (Fig.~\ref{fig:nv}(a)) is certainly very similar to the zonal-wind pattern of EOF1 (Fig.~\ref{fig:test3}(a)), with the exception of small differences in the stratosphere around $40^\mathrm{o}$. The temperature patterns, however, are in general different except in the midlatitude around $30^\mathrm{o}-50^\mathrm{o}$. 

The above calculations of singular vectors are subject to uncertainties in the relative weights used for the different variables (e.g., zonal-wind versus temperature), or in other words, the choice of norm \citep{kuang2004norm}. However, below we show that the difference between the neutral vector of ${\hat{\pmb{\mathsf{M}}}}$ and EOF1 in our results is mostly due to fact that the stochastic eddy forcing is not isotropic and uncorrelated and therefore does not satisfy the above condition. This is demonstrated by calculating the EOF1 of $\overline{\mathrm{\mathbf{z}}}$ obtained from a long integration of the stochastic linear equation 
\begin{eqnarray}
\dot{\overline{\mathrm{\mathbf{z}}}} =   \hat{\pmb{\mathsf{M}}} \, \overline{\mathrm{\mathbf{z}}} +\boldsymbol{\zeta},
\end{eqnarray}
where $\boldsymbol{\zeta}(t)$ is Gaussian white noise, $\langle \boldsymbol{\zeta} \boldsymbol{\zeta}^\dag  \rangle = \pmb{\mathsf{I}}$ (see section~5 of Part~2 for details). For this case the conditions on the forcing are satisfied and the EOF1 is almost identical to the neutral vector of $\hat{\pmb{\mathsf{M}}}$ in both zonal-wind and temperature patterns (see Fig.~S4). 

The forcing needed to produce the neutral vector (Figs.~\ref{fig:nv}(c)-\ref{fig:nv}(d)) includes a strong heating in the subtropical upper troposphere and a dipolar torquing pattern that is centered around $40^\mathrm{o}$ and is broader (in latitude) compared to the dipolar pattern of the neutral vector's zonal-wind (Figs.~\ref{fig:nv}(a)).       

We further highlight that despite various simplifications in the idealized GCM used here, the pattern of the Annular Mode in this model (i.e., the EOF1 shown in Fig.~\ref{fig:test3}(a)-\ref{fig:test3}(b)) resembles the observed patterns of the Northern Annular Mode (NAM) \citep[][Fig.~2]{thompson2015baroclinic} and the Southern Annular Mode (SAM) \citep[][Fig.~2]{thompson2014barotropic} particularly for zonal-winds (also note that the details of EOF calculation in the current study are different from those in the two aforementioned papers; here the EOF is calculated from extended EOF analysis of unweighted $(\overline{u},\overline{T})$). These similarities and the above discussion on the connection between neutral vector and EOF1 suggest that it is plausible that the neutral vector of the extratropical atmospheric circulation in more complex GCMs and the real atmosphere resembles the patterns in Fig.~\ref{fig:nv} particularly for the zonal-wind, which would explain the ubiquity of Annular Mode-like patterns in the midlatitude response to external forcings in full-physics GCM \citep[e.g.,][]{peings2014response,deser2015role}.                   

\begin{figure*}[t]
\centerline{\includegraphics[width=1\textwidth]{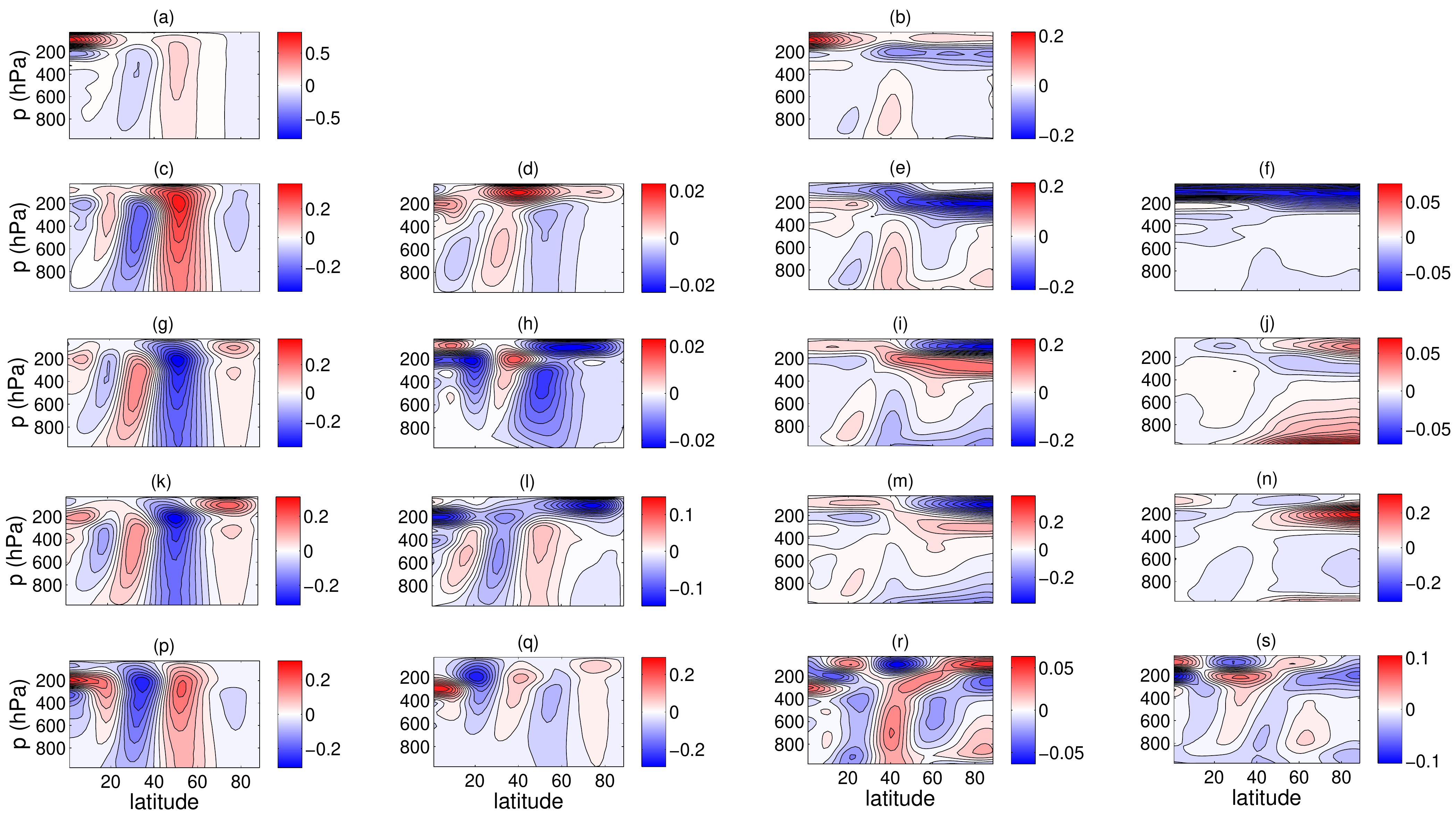}}
\caption{Selected eigenvectors of the slowest decaying modes of $\tilde{\pmb{\mathsf{M}}}$. First few slowest decaying modes with clearly distinct eigenvectors or eigenvalues are selected (only one mode is shown for complex pairs). Two left (right) columns show zonal-wind (temperature) and the first and third (second and fourth) columns from the left show the real (imaginary) part of the eigenvector. (a)-(b): slowest decaying eigenvector with eigenvalue $\lambda=-0.017$~day$^{-1}$. (c)-(f): second slowest decaying eigenvector with $\lambda=-0.031-0.004\, \mathrm{i}$~day$^{-1}$. (g)-(j): fourth slowest decaying eigenvector with $\lambda=-0.050+0.001\, \mathrm{i}$~day$^{-1}$. The third slowest decaying mode with $\lambda=-0.041$~day$^{-1}$ is similar to (g) and (i)  and is not shown. (k)-(n): fifth slowest decaying eigenvector with $\lambda=-0.059-0.019\, \mathrm{i}$~day$^{-1}$. (p)-(s): sixth slowest decaying eigenvector with $\lambda=-0.060-0.106\, \mathrm{i}$~day$^{-1}$.}
\label{fig:S3}
\end{figure*}

We also briefly discuss the eigenmodes of ${\tilde{\pmb{\mathsf{M}}}}$ (the eigenmodes of ${\hat{\pmb{\mathsf{M}}}}$ are similar with slightly reduced eigenvalues due to the filtering (\ref{eq:M2})). In Fig.~S1 we show the eigenvalues of $\tilde{\pmb{\mathsf{M}}}$ with timescales longer than $1$~day, which are all decaying and the slowest decaying modes have timescales on the order of tens of days. Selected eigenvectors of the slowest decaying modes are shown in Fig.~\ref{fig:S3}. The zonal-wind and temperature of the slowest decaying eigenvector are mostly confined to the stratosphere, although there is a weak Annular Mode signature in the troposphere. The decaying timescale is $\sim 59$~days, which is comparable to the imposed $40$-day Newtonian relaxation time. The next few slowest decaying modes have timescales $\sim 30$~days or shorter and zonal-wind patterns with strong Annular Mode signatures. 
%It is worth mentioning that the Annular Mode of the model has a decorrelation time of $\sim 45$~days.                       

\section{Applications of the LRF and EFM} \label{sec:app}
In this section we briefly discuss some of the potential applications of $\hat{\pmb{\mathsf{M}}}$ and $\hat{\pmb{\mathsf{E}}}$ which can be categorized as i) forcing a specified mean-flow for hypothesis-testing, ii) isolating and quantifying eddy-feedbacks, and iii) examining and evaluating more generally-applicable methods.

One difficulty in fully understanding the dynamics of many complex atmospheric phenomena even using extensive GCM experiments is that several changes in the large-scale circulation, such as the speed and latitude of the jet-streams and static stability, might happen simultaneously in response to a forcing or varying a physical parameter, which obscure understanding the direction of causation and the individual contribution of each change. This problem exists even in idealized GCMs such as the one used here, and common but imperfect remedies include nudging and/or tuning the forcing to control the changes in the large-scale circulation \citep{kidston11,simpson2013southern,garfinkel2013effect}. The LRF can be used to accurately calculate time-invariant forcings needed to generate a desired mean-flow (as done in Tests 2-3) and use the well-controlled experiments to disentangle the influences of different changes in the mean-flow on the phenomenon under study.

For example, in \citet{pedram15}, we used the LRF to examine causality in the relationship between the negative phase of NAM and increased atmospheric blocking, which is seen in observational data and GCM simulations, including the idealized GCM used here. To test whether the mean-state of the negative phase of NAM (i.e., the patterns in Figs.~\ref{fig:test3}(a)-(b) with opposite signs) causes more blocking, $\hat{\pmb{\mathsf{M}}}$ was used to force this mean-state with various amplitudes (similar to Test~3), and it was found that blocking decreases as the amplitudes increases. Results suggest that the observed blocking-NAM relationship is a correlation which does not imply that the mean-state of the negative phase of NAM causes more blocking. These findings have important implications for the ongoing debate on the linkage between Arctic Amplification and the midlatitude weather extremes \citep[see, e.g.,][]{pedram14,barnes2015}.

As discussed in section~\ref{sec:intro}, isolating and quantifying the eddy-feedbacks is another difficulty in developing a full mechanistic understanding of problems in which eddy-mean flow interaction plays an important role; examples of such problems include the dynamics of the Annular Modes and their overestimated persistence in GCMs \citep{feldstein1998atmospheric,robinson2000baroclinic,lorenz2001eddy,simpson2013southern,nie2014quantifying}, poleward shift of the midlatitude jets under climate change \citep{chen2007phase,kidston11,lorenz2014understanding} and troposphere-stratosphere coupling \citep{kushner2004stratosphere,song2004dynamical,domeisen2013role}. The EFM $\hat{\pmb{\mathsf{E}}}$ and similar matrices for eddy-phase speed and eddy length-scale can be used to isolate and quantify the response of eddies to a given change in the mean-flow, which combined with well-controlled GCM experiments made possible using $\hat{\pmb{\mathsf{M}}}$, can help with developing a deeper dynamical understanding of these problems.

A limitation of the approach presented here is that to study various problems involving different physical processes, LRF and EFM should be recalculated for GCMs that represent these processes. However, it should be highlighted that the dry dynamical core with Held-Suarez physics is a widely-used GCM that provides a dynamical framework to study the role of dry processes in various complex problems. Thus the calculated LRF and EFM can be used to study a variety of problems (some mentioned above). Furthermore, the approach presented in this paper can be applied to construct the LRFs and EFMs for more complex GCMs and for zonally-asymmetric forcings/responses. For more complex GCMs, the main difficulty will be likely the computational cost associated with the larger size of the state-vector (i.e., number of variables) and achieving a reasonable signal-to-noise ratio in the linear regime. Note that for zonally-asymmetric forcing/responses, it might be better to use basis functions that are vertically in grid-space (used here in both directions) but horizontally in spectral-space (i.e., use low-wavenumber spherical harmonics).                                  

Additionally, the LRF and EFM constructed here can be used to evaluate, complement, and potentially improve generally-applicable methods that are currently employed to construct LRFs and quantify eddy-feedbacks in the outputs of idealized and comprehensive GCMs . Such methods include FDT, to construct LRFs, and lag-regression \citep{lorenz2001eddy,simpson2013southernII} and finite-amplitude wave-activity \citep{nakamura2010finite,nie2014quantifying}, to diagnose eddy feedbacks. As an example, in Part~2 of this study, we use $\hat{\pmb{\mathsf{M}}}$ to investigate why the LRF constructed using the FDT performs poorly in some cases.   

\section{Summary} \label{sec:sum}
In Part~1 of this study, we have calculated the LRF, $\hat{\pmb{\mathsf{M}}}$, and EFM, $\hat{\pmb{\mathsf{E}}}$, for an idealized dry GCM using Green's functions and described the procedure in details (sections~\ref{sec:eqs}-\ref{sec:const}). Several tests in section~\ref{sec:test} show that the LRF accurately predicts the mean-response to imposed thermal/mechanical forcings and vice versa and the EFM accurately predicts changes in eddy fluxes in response to a change in the mean-flow. The spectral analysis of the LRF (section~\ref{sec:nv}) reveals that the model's most excitable mode, i.e., the neutral vector, strongly resembles its Annular Mode, in particular for zonal-wind, which suggests that the Annular Mode is truly a dynamical mode of the system and explains the ubiquity of Annular Mode-like responses to external forcings. In section~\ref{sec:app} we discuss the potential applications of $\hat{\pmb{\mathsf{M}}}$ and $\hat{\pmb{\mathsf{E}}}$ which include i) forcing a specified mean-flow for hypothesis-testing, ii) isolating and quantifying eddy-feedbacks in eddy-mean flow interaction problems, and iii) examining and evaluating more generally-applicable methods such as the FDT. 

In Part~2 of this study \citep{pedram16}, we calculate, for the same idealized GCM, the LRF using the FDT and investigate the source(s) of its poor performance for some tests by employing the LRF calculated using the Green's functions in Part~1. We show that for non-normal operators, dimension-reduction by projecting the data onto the leading EOFs, which is commonly used to construct LRFs from the FDT, can significantly degrade the accuracy. 

%%%%%%%%%%%%%%%%%%%%%%%%%%%%%%%%%%%%%%%%%%%%%%%%%%%%%%%%%%%%%%%%%%%%%
% FIGURES---PLACE AT END OF DOCUMENT
%%%%%%%%%%%%%%%%%%%%%%%%%%%%%%%%%%%%%%%%%%%%%%%%%%%%%%%%%%%%%%%%%%%%%

%

%%%%%%%%%%%%%%%%%%%%%%%%%%%%%%%%%%%%%%%%%%%%%%%%%%%%%%%%%%%%%%%%%%%%%
% TABLES---PLACE AT END OF DOCUMENT
%%%%%%%%%%%%%%%%%%%%%%%%%%%%%%%%%%%%%%%%%%%%%%%%%%%%%%%%%%%%%%%%%%%%%
%\begin{table}[h]
%\caption{This is a sample table caption and table layout.  
%Table from Lorenz (1963).}\label{t1}
%\begin{center}
%\begin{tabular}{ccccrrcrc}
%\topline
%$N$ & $X$ & $Y$ & $Z$\\
%\midline
% 0000 & 0000 & 0010 & 0000 \\
% 0005 & 0004 & 0012 & 0000 \\
% 0010 & 0009 & 0020 & 0000 \\
% 0015 & 0016 & 0036 & 0002 \\
% 0020 & 0030 & 0066 & 0007 \\
% 0025 & 0054 & 0115 & 0024 \\
%\botline
%\end{tabular}
%\end{center}
%\end{table}
%%

%%%%%%%%%%%%%%%%%%%%%%%%%%%%%%%%%%%%%%%%%%%%%%%%%%%%%%%%%%%%%%%%%%%%%
% ACKNOWLEDGMENTS
%%%%%%%%%%%%%%%%%%%%%%%%%%%%%%%%%%%%%%%%%%%%%%%%%%%%%%%%%%%%%%%%%%%%%
\acknowledgments
We thank Ashkan Borna, Gang Chen, Brian Farrell, Nick Lutsko, Ding Ma, and Saba Pasha for fruitful discussions; Elizabeth Barnes, Packard Chan, Nick Lutsko, Marie McGraw, and Marty Singh for useful comments on the manuscript; and Chris Walker for help with the GCM runs at the initial stage of this study. We are grateful to two anonymous reviewers for insightful feedbacks. This work was supported by a Ziff Environmental Fellowship from the Harvard University Center for the Environment to P.H. and NSF grant AGS-1062016 to Z.K. The simulations were run on Harvard Odyssey cluster. 

%%%%%%%%%%%%%%%%%%%%%%%%%%%%%%%%%%%%%%%%%%%%%%%%%%%%%%%%%%%%%%%%%%%%%
% APPENDIXES
%%%%%%%%%%%%%%%%%%%%%%%%%%%%%%%%%%%%%%%%%%%%%%%%%%%%%%%%%%%%%%%%%%%%%

%% If only one appendix, use
\appendix[A]
\label{app:A}
We start from the zonally-averaged zonal-momentum and temperature equations, linearized around the mean-flow $(\langle \overline{U} \rangle,\langle \overline{V} \rangle,\langle \overline{\Omega} \rangle,\langle \overline{\Theta} \rangle )$ while the eddy fluxes are retained:
\begin{eqnarray}
\frac{\partial \overline{u}}{\partial t} &=& - \frac{\langle \overline{V} \rangle}{a \cos{\mu}} \frac{\partial (\overline{u}\cos{\mu})}{\partial \mu} - \langle \overline{\Omega} \rangle \frac{\partial \overline{u}}{\partial p} + f \overline{v} -  
\nonumber \\
&& \frac{\overline{v}}{a \cos{\mu}} \frac{\partial (\langle \overline{U} \rangle \cos{\mu})}{\partial \mu} - \overline{\omega} \frac{\partial \langle \overline{U} \rangle }{\partial p} +  \nonumber \\
&& E_u + k_u \overline{u} + \overline{f}_u 
\label{eq:lrf1Au} \\
&& \nonumber \\
\frac{\partial \overline{T}}{\partial t} &=& - \frac{\langle \overline{V} \rangle}{a} \frac{\partial \overline{T}}{\partial \mu} - \langle \overline{\Omega} \rangle \left[\frac{\partial \overline{T}}{\partial p}-\frac{\kappa \overline{T}}{p} \right]-  
\nonumber \\
&& \frac{\overline{v}}{a} \frac{\partial \langle \overline{\Theta} \rangle}{\partial \mu} - \overline{\omega} \left[\frac{\partial \langle \overline{\Theta} \rangle }{\partial p} - \frac{\kappa \langle \overline{\Theta} \rangle}{p}\right]+  \nonumber \\
&& E_T + k_T \overline{T} + \overline{f}_T 
\label{eq:lrf1At}
\end{eqnarray}
where $f$ is the Coriolis frequency (not to be confused with the external forcing); $a$ is the radius of Earth, $\kappa=(c_p-c_v)/c_p=2/7$ ($c_p$ and $c_v$ are the specific heats); $E_u$ and $E_T$ are, respectively, the divergence of eddy momentum and heat fluxes; and $k_u$ and $k_T$ are, respectively, the Rayleigh drag and Newtonian cooling damping rates of the Held-Suarez physics. 

Using the continuity equation, $(\overline{v},\overline{\omega})$ can be represented using streamfunction $\chi$ as $(\partial \chi/\partial p,-(\partial (\chi \cos{\mu})/\partial \mu)/(a \cos{\mu}))$, and using assumption~2 (i.e., quasi-equilibrium; see section~\ref{sec:eqs}), $(E_u,E_T)$ can be represented as a linear function of $(\overline{u},\overline{T})$ (although we emphasize that this linear function is unknown). Then Eqs.~(\ref{eq:lrf1Au}) and (\ref{eq:lrf1At}) can be written as      
\begin{eqnarray}
\dot{\overline{\mathrm{\mathbf{y}}}} = \hat{\pmb{\mathsf{L}}} \, \overline{\mathrm{\mathbf{y}}} + \pmb{\mathsf{H}} \, \chi +       \overline{\mathrm{\mathbf{f}}}
\label{eq:lrf2A}
\end{eqnarray}
where $\hat{\pmb{\mathsf{L}}} \, \overline{\mathrm{\mathbf{y}}}$ represents the terms on the first and third lines of the right-hand side of Eqs.~(\ref{eq:lrf1Au}) and (\ref{eq:lrf1At}), except for the external forcing terms which are represented by $\overline{\mathrm{\mathbf{f}}}$. Operator $\pmb{\mathsf{H}}$ is a function of $\langle \overline{U} \rangle$ and $\langle \overline{\Theta} \rangle$ and the second term on the right-hand side of (\ref{eq:lrf2A}) represents meridional advection of these quantities by $\chi$ (i.e., the second lines on the right-hand side of (\ref{eq:lrf1Au}) and (\ref{eq:lrf1At})). 

If $(\overline{u},\overline{T})$ are in gradient-wind balance and/or if the tendencies on the left-hand side of Eqs.~(\ref{eq:lrf1Au}) and (\ref{eq:lrf1At}) are negligible compared to the dominant terms on the right-hand sides (which is reasonable given assumption~2, i.e., quasi-equilibrium), then the right-hand sides of (\ref{eq:lrf1Au}) and (\ref{eq:lrf1At}) can be combined to find a diagnostic equation for $\chi$:
\begin{eqnarray}
\chi =  \pmb{\mathsf{G}} \, \overline{\mathrm{\mathbf{y}}} + \pmb{\mathsf{K}} \, \overline{\mathrm{\mathbf{f}}}.
\label{eq:lrf2B}
\end{eqnarray}          
The first term on the right-hand side represents the component of $\chi$ due to $\overline{\mathrm{\mathbf{y}}}$, and the second term represents the component of $\chi$ due to the part of $\overline{\mathrm{\mathbf{f}}}$ that is not in gradient-wind balance (if any). The latter is related to the Eliassen's balanced vortex problem \citep{eliassen1951slow}. According to Eq.~(\ref{eq:lrf2B}), given the above conditions, the meridional circulation is in quasi-equilibrium with $(\overline{u},\overline{T},\overline{f})$.       

Substituting (\ref{eq:lrf2B}) in (\ref{eq:lrf2A}) and rearranging the terms yield
\begin{eqnarray}
\dot{\overline{\mathrm{\mathbf{y}}}} = (\hat{\pmb{\mathsf{L}}}+\pmb{\mathsf{H}}\pmb{\mathsf{G}}) \, \overline{\mathrm{\mathbf{y}}} + ({\pmb{\mathsf{I}}}+\pmb{\mathsf{H}}\pmb{\mathsf{K}}) \,  \overline{\mathrm{\mathbf{f}}}
\label{eq:lrf2C}
\end{eqnarray}
Equation~(\ref{eq:lrf3}) follows with defining ${\pmb{\mathsf{M}}} \equiv \hat{\pmb{\mathsf{L}}}+\pmb{\mathsf{H}}\pmb{\mathsf{G}}$ and ${\pmb{\mathsf{B}}} \equiv {\pmb{\mathsf{I}}}+\pmb{\mathsf{H}}\pmb{\mathsf{K}}$. ${\pmb{\mathsf{B}}}$ only depends on $\langle \overline{U} \rangle$ and $\langle \overline{\Theta} \rangle$ and if needed, can be analytically calculated for the Held-Suarez setup.

It should be mentioned that with additional assumptions the state-vector could be reduced to only one variable. For example, \citet{ring08} only used $\overline{u}$ as the state-vector, which requires:
\begin{itemize}
\item Assuming the gradient-wind balance and replacing $\partial \overline{T}/\partial \mu$ with $\partial \overline{u}/\partial p$ (multiplied with the appropriate coefficients) in the formulation of $\chi$,
\item Assuming that the eddy fluxes are insensitive to changes in the static stability, 
\item Assuming that the second term on the right-hand side of Eq.~(\ref{eq:lrf2B}) can be neglected,
\end{itemize}  
in addition to assumptions~1 and 2 in the current study. As noted by \citet{ring08}, the second assumption can be particularly problematic; see their paper for further discussions of these assumptions. With these assumptions, $\chi$ can be represented only as a function of $\overline{u}$ and the state-vector reduces to one variable. 

Also note that although we use $(\overline{u},\overline{T})$ and \citet{ring08} used $\overline{u}$, the procedures of the state-vector reduction are very similar and the reader is encouraged to consult the Appendix~A in \citet{ring08} as they derive/present most of the operators and equations mentioned above in details.

\appendix[B]
\label{app:criteria}
Choosing the appropriate forcing amplitude $\bar{f}_o$ in each trial can be challenging, which is a common issue in problems involving LRFs. As noted by \citet{leith1975climate} and \citet{ring08}, obtaining statistically robust results (i.e., large signal-to-noise ratios) and maintaining the linear regime (assumption 1) at the same time is difficult because the former requires strong forcings while the latter requires small forcings. What further complicates the problem for the current study (and likely for other studies involving the large-scale circulation) is that the response of the extratropics to external forcings projects strongly onto the leading pattern of internal variability (i.e., the Annular Modes), which makes it difficult to distinguish the signal from the noise. The projection of forced responses onto patterns of internal variability is seen in GCMs with various degree of complexity    \citep[e.g.,][]{ring2007forced,butler2010steady,deser2004effects} and in observations \citep{corti1999signature,thompson2002interpretation} and has been discussed in the context of FDT \citep[e.g.,][]{shepherd2014} and neutral vectors \citep{palmer99,goodman2002using} (also see section~\ref{sec:nv}). 

\begin{figure*}[t]
\renewcommand\thefigure{B1}
\centerline{\includegraphics[width=1\textwidth]{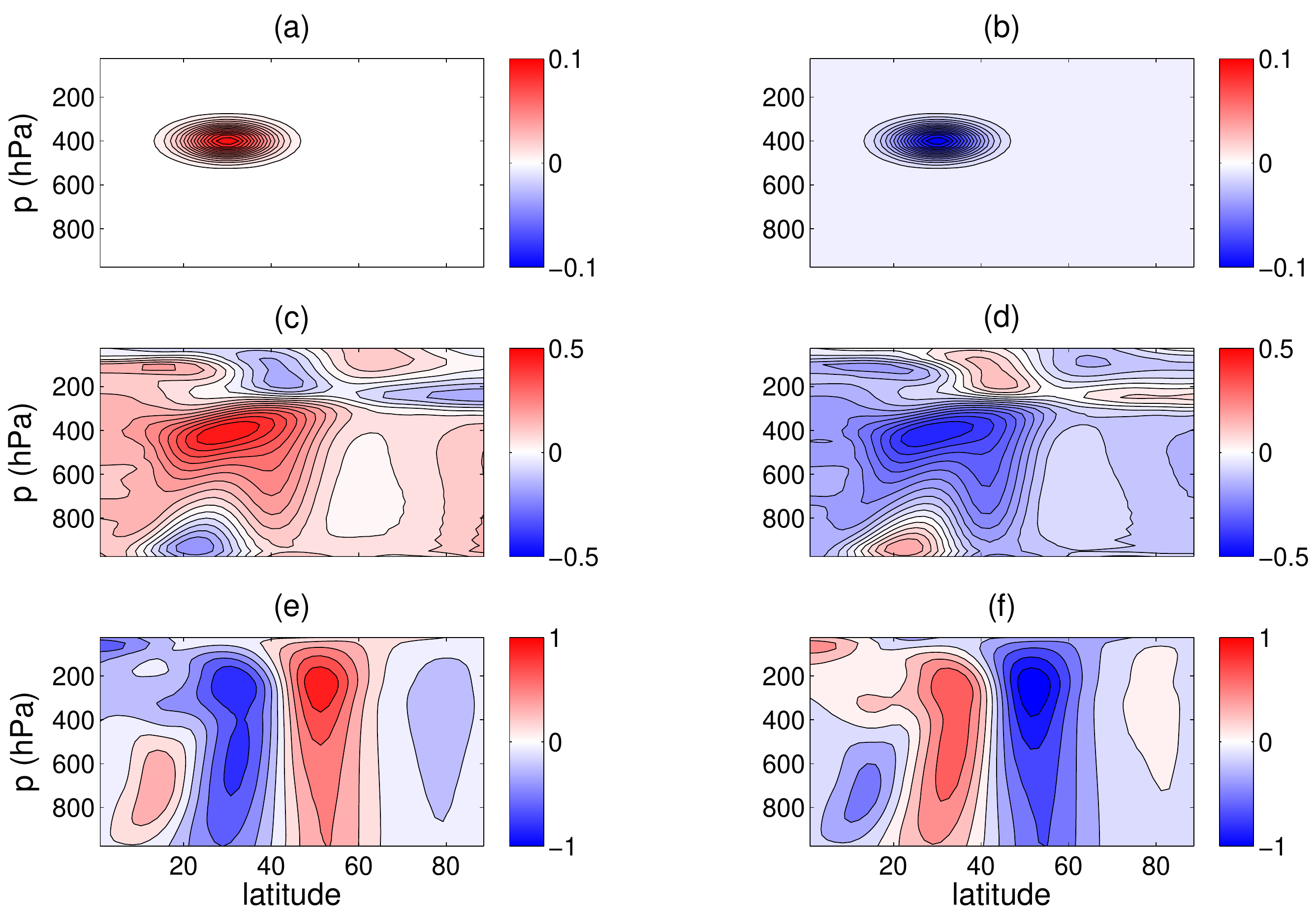}}
\caption{Examples of results with imposed heating (left) and cooling (right) at $\mu_o=30^\mathrm{o}$ and $p_o=400$~hPa with the amplitude of $0.1$~K~day$^{-1}$. Heating (a) and cooling (b) profiles. $\langle \overline{T} \rangle = \langle \overline{\Theta} \rangle_+ - \langle \overline{\Theta} \rangle_c$ in K in response to heating (c) and cooling (d). $\langle \overline{u}  \rangle = \langle \overline{U} \rangle_+ - \langle \overline{U} \rangle_c$ in m~s$^{-1}$ in response to heating (e) and cooling (f). The relative errors ($e$) and signal-to-noise ratios ($\mathrm{SNR}$), defined in Eqs.~(\ref{eq:e}) and (\ref{eq:snr}), are $e_\infty(\langle \overline{\Theta} \rangle)=14\%$, $e_2(\langle \overline{\Theta} \rangle)=4\%$, $e_\infty(\langle \overline{U} \rangle)=3\%$, $e_2(\langle \overline{U} \rangle)=3\%$; $\mathrm{SNR}(\langle \overline{T} \rangle_+)=17$ and $\mathrm{SNR}(\langle \overline{T} \rangle_-)=5$; $\mathrm{SNR}(\langle \overline{U} \rangle_+)=12$ and $\mathrm{SNR}(\langle \overline{U} \rangle_-)=3$.}
\label{fig:1}
\end{figure*}

We choose $\bar{f}_o$ by trial-and-error, at least for the early trials, where we explore selected $(\mu_o,p_o)$ for $\mu_o=0^\mathrm{o},30^\mathrm{o},60^\mathrm{o},90^\mathrm{o}$ and $p_o=300,600,900$~hPa to find $\bar{f}_o$ that produces a reasonable signal-to-noise ratio within the linear regime for each $(\mu_o,p_o)$ and each variable ($U$ or $\Theta$). Knowing acceptable $\bar{f}_o$ of these selected trials, $\bar{f}_o$ for other trials can be reasonably guessed (in some cases further trial-and-error is needed). To determine whether a forcing amplitude is acceptable, we employ three criteria (one qualitative and two quantitative) to evaluate the signal-to-noise ratio and linearity using the inter-hemispheric asymmetry of each trial and differences between the trials forced with $\pm \bar{f}_o$:  
\begin{enumerate}
\item We visually inspect the hemispheric-symmetry in each of the following four patterns $\langle \overline{U} \rangle_{\pm}$ and $\langle \overline{\Theta} \rangle_{\pm}$. Large asymmetries indicate (qualitatively) small signal-to-noise ratios. We also visually compare the patterns of hemispherically-averaged $\langle \overline{U} \rangle_{+}$ with $\langle \overline{U} \rangle_{-}$, and $\langle \overline{\Theta} \rangle_{+}$ with $\langle \overline{\Theta} \rangle_{-}$. Large differences indicate small signal-to-noise ratios and/or nonlinearity, both of which are undesirable. We require these differences and asymmetries to be reasonably small.               
\item Two measures of relative error $e$ are calculated: 
\begin{eqnarray}
\!\!\!\!\!\!\!\!\!\!\!\! e(a) \!\!\!\!\!\! &\equiv& \!\!\!\!\!\! \frac{|\|a_+-a_c\|-\|a_{-}-a_c\||}{(\|a_+-a_c\|+\|a_{-}-a_c\|)/2}  \times 100
\label{eq:e}
\end{eqnarray}
where $a$ is hemispherically-averaged $\langle \overline{U} \rangle$ or $\langle \overline{\Theta}\rangle$ and the norms are either 
\begin{eqnarray}
\|a\|_\infty &\equiv& \mathrm{max}(|a|) \label{eq:max}\\ 
\|a\|_2 &\equiv& \sqrt{\Sigma \, a^2} \label{eq:l2}
\end{eqnarray}
where max and $\Sigma$ are over the latitude-pressure domain. Large $e_2$ or $e_\infty$ show small signal-to-noise ratios and/or nonlinearity. We require $e_2$ and $e_\infty$ to be $\leq 20\%$ for both variables.     
\item A measure of the signal-to-noise ratio is defined as 
\begin{eqnarray}
\!\!\!\!\!\!\!\!\!\!\!\! \mathrm{SNR}(a_\pm) \!\!\!\!\!\! &\equiv& \!\!\!\!\!\! \frac{\|(a_\pm-a_c)_\mathrm{NH}+(a_\pm-a_c)_\mathrm{SH}\|_\infty}{\|(a_\pm)_\mathrm{NH}-(a_\pm)_\mathrm{SH}\|_\infty}
\label{eq:snr}
\end{eqnarray}
where NH~(SH) refer to the Northern (Southern) hemisphere, and $a$ is either $\langle \overline{U} \rangle$ or $\langle \overline{\Theta}\rangle$. We require $\mathrm{SNR}$ to be at least $3$ for both variables although the number is larger than $5$ in most accepted trials.   
\end{enumerate}
A forcing amplitude $\bar{f}_o$ is acceptable for each basis function and used in the calculation of $\tilde{\pmb{\mathsf{M}}}$ if all the three criteria are satisfied. The accepted values of $\bar{f}_o$ are shown in Tables~S1-S2 (Supplemental Material). As a rule of thumb, stronger forcings are needed poleward and toward the surface, and for $U$ compared to $\Theta$. In Fig.~\ref{fig:1} we show an example of the typical results from a pair of trials that are accepted and used in the calculation of the LRF and EFM.                   

Finally, it should be mentioned that external torque at the highest pressure level ($\sim 9$~hPa) is found to result in unrealistically large responses in the tropical stratosphere (see \citet{scott1998internal} for a discussion). To avoid this problem, the basis functions of the zonal-wind for $p_o=100$~hPa are set to zero at the highest pressure level and small-amplitude forcings are applied. 

%% If more than one appendix, use \appendix[<letter>], e.g.,
%\appendix[C]

%%%%%%%%%%%%%%%%%%%%%%%%%%%%%%%%%%%
%APPENDIX FIGURE AND TABLE EXAMPLES--PLACE AT END OF DOCUMENT
%%%%%%%%%%%%%%%%%%%%%%%%%%%%%%%%%%%
%
%\begin{table}
%\appendcaption{A1}{Here is the appendix table caption.}
%\centering
%\begin{tabular}{ccc}
%a&b&c\\
%d&e&f
%\end{tabular}
%\end{table}
%
%\begin{figure}
%\centerline{(illustration here)}
%\appendcaption{A1}{Here is the appendix figure caption.}
%\end{figure}
%

%\appendix[B]
%\appendixtitle{File Structure of the AMS \LaTeX\ Package}

%\subsection{AMS \LaTeX\ files}

%%%%%%%%%%%%%%%%%%%%%%%%%%%%%%%%%%%%%%%%%%%%%%%%%%%%%%%%%%%%%%%%%%%%%
% REFERENCES
%%%%%%%%%%%%%%%%%%%%%%%%%%%%%%%%%%%%%%%%%%%%%%%%%%%%%%%%%%%%%%%%%%%%%

\bibliographystyle{ametsoc2014}
\setlength{\bibsep}{2pt plus 0.75ex}
\bibliography{LRFp1_v1}

%%%%%%%%%%%%%%%%%%%%%%%%%%%%%%%%%%%%%%%%%%%%%%%%%%%%%%%%%%%%%%%%%%%%%
% TABLES
%%%%%%%%%%%%%%%%%%%%%%%%%%%%%%%%%%%%%%%%%%%%%%%%%%%%%%%%%%%%%%%%%%%%%

%%%%%%%%%%%%%%%%%%%%%%%%%%%%%%%%%%%%%%%%%%%%%%%%%%%%%%%%%%%%%%%%%%%%%
% FIGURES
%%%%%%%%%%%%%%%%%%%%%%%%%%%%%%%%%%%%%%%%%%%%%%%%%%%%%%%%%%%%%%%%%%%%%

\end{document}